\documentclass[aps,preprintnumbers,amsmath,amssymb,prd,superscriptaddress,longbibliography,twocolumn]{revtex4-1}
\usepackage{dcolumn}
\usepackage{color}
\usepackage[caption=false]{subfig}
\usepackage{feynmp}
\usepackage{feynmp-auto}
\usepackage{systeme,mathtools}
\usepackage{easyReview}

\def\comment#1{}

\newcommand{\nc}{\newcommand}
\nc{\beq}{\begin{eqnarray}}
\nc{\eeq}{\end{eqnarray}}
\nc{\scs}{\scriptstyle}
\nc{\setval}{\fmfset{wiggly_len}{3mm} \fmfset{arrow_len}{1.5mm}
	\fmfset{arrow_ang}{13} \fmfset{dash_len}{1.5mm}\fmfpen{0.125mm}
	\fmfset{dot_size}{2thick}}

\usepackage{bm,latexsym,mathrsfs,enumerate,color}
\usepackage[mathcal]{euscript}
\usepackage[breaklinks=true,unicode=true,urlcolor = blue,colorlinks = true,citecolor = blue,linkcolor = blue]{hyperref}
\usepackage{graphicx}
\usepackage{todonotes}
\usepackage{wrapfig}
\usepackage{easyReview}
\usepackage{mathbbol}
\usepackage{stmaryrd}

\renewcommand{\vec}[1]{\bm{#1}}

\def\slashchar#1{\setbox0=\hbox{$#1$}           
	\dimen0=\wd0                                 
	\setbox1=\hbox{/} \dimen1=\wd1               
	\ifdim\dimen0>\dimen1                        
	\rlap{\hbox to \dimen0{\hfil/\hfil}}      
	#1                                        
	\else                                        
	\rlap{\hbox to \dimen1{\hfil$#1$\hfil}}   
	/                                         
	\fi}                                         %

\DeclareMathAlphabet\mathbfcal{OMS}{cmsy}{b}{n}
\DeclareSymbolFontAlphabet{\amsmathbb}{AMSb}%

\def\sigmab{{\mbox{\boldmath $\sigma$}}}

\begin{document}
	
\title{Bosonization duality in 2+1 dimensions and critical current correlation functions in Chern-Simons $U(1)\times U(1)$ Abelian Higgs model}

\author{Vira Shyta}
\affiliation{Institute for Theoretical Solid State Physics, IFW Dresden, Helmholtzstr. 20, 01069 Dresden, Germany}

\author{Flavio S. Nogueira}
\affiliation{Institute for Theoretical Solid State Physics, IFW Dresden, Helmholtzstr. 20, 01069 Dresden, Germany}

\author{Jeroen van den Brink}
\affiliation{Institute for Theoretical Solid State Physics, IFW Dresden, Helmholtzstr. 20, 01069 Dresden, Germany}
\affiliation{Institute for Theoretical Physics and W\"urzburg-Dresden Cluster of Excellence ct.qmat, TU Dresden, 01069 Dresden, Germany}

\begin{abstract}
	While the phase structure of the $U(1)\times U(1)$-symmetric Higgs theory is still under debate, a version of this theory with an additional Chern-Simons term was recently shown to  undergo a second-order phase transition [V. Shyta, J. van den Brink, and F. S. Nogueira, Phys. Rev. Lett. 127, 045701 (2021)]. This theory is dual to a topological field theory of massless fermions featuring two gauge fields. 
	Here we elaborate on several aspects of this duality, focusing on the critical current correlators and on the nature of the critical point as reflected by the bosonization duality. The current correlators associated to the $U(1)\times U(1)$ symmetry and the topological current are shown to coincide up to a universal prefactor, which we find to be the same for both $U(1)$ and $U(1)\times U(1)$ topological Higgs theories. 
	The established duality offers in addition another way to substantiate the claim about the existence of a critical point in the bosonic Chern-Simons $U(1)\times U(1)$ Higgs model: a Schwinger-Dyson analysis of the fermionic dual model shows that no dynamical mass generation occurs. The same cannot be said for the theory without the Chern-Simons term in the action.   
\end{abstract}
\maketitle

\section{Introduction}

The study of bosonization in dimensions higher than 1+1 has been around for a long time \cite{LUSCHER1989557,Polyakov-1988,FRADKIN1994253}. While bosonization in 1+1 dimension is a well developed field \cite{giamarchi2003quantum,zinn2002quantum}, this does not hold for the (2+1)-dimensional case, in spite of the intense recent activity in the field \cite{SEIBERG2016395,Aharony-JHEP-2016,Karch_PhysRevX.6.031043,Senthil_PhysRevX.7.031051,metlitski2015sduality,NASTASE2018145,Raghu_PhysRevLett.120.016602,turker2020bosonization,Shyta_PhysRevLett.127.045701}. Typically, in 1+1 dimensions scalar fields are expressed {\it locally} in terms of fermion bilinears \cite{Coleman_PhysRevD.11.2088,Witten-nonabelbos}. One such fermion bilinear is the current $j_\mu=i\bar{\psi}\gamma_\mu\psi$, but other local forms are also possible. A well known example is the bosonization of the Thirring model, where the fermion current is expressed as $j_\mu\sim \epsilon_{\mu\nu}\partial_\nu\varphi$ \cite{Coleman_PhysRevD.11.2088}. Since $j_\mu$ is a vector, going one dimension higher implies necessarily that bosonization in 2+1 dimensions should involve a gauge field, $a_\mu$, therefore, one can guess, $j_\mu\sim \epsilon_{\mu \nu \lambda}\partial_\nu a_\lambda$. Of course, this intuitive picture has to be substantiated by actual calculations and it becomes immediately clear that unlike the (1+1)-dimensional case, more fields have to be involved, so the gauge field has to be minimally coupled to a scalar field in order to match the symmetries of fermionic and bosonic theories \cite{SEIBERG2016395}. As a consequence, in the case of $U(1)$ symmetries in 2+1 dimensions the bosonization manifests itself as a kind of particle-vortex duality \cite{Peskin1978,THOMAS1978513,Dasgupta-Halperin_PhysRevLett.47.1556,kleinert1982disorder}. One important feature of dualities is that they provide mappings between strongly and weakly coupled theories, thus presenting a non-perturbative tool to analyze quantum field theories. Dualities between bosonic $U(1)$-symmetric theories on the lattice have been established exactly in several dimensions \cite{BANKS1977493,Jose_PhysRevB.16.1217,Savit_RevModPhys.52.453,Peskin1978,CARDY198217,CARDY1980369}. Since bosonization dualities in 2+1 dimensions involve Chern-Simons (CS) terms \cite{SEIBERG2016395,Karch_PhysRevX.6.031043}, it is considerably more difficult to perform the procedure on the lattice, although it has been considered in some cases more rigorously for gapped fermions \cite{Raghu_PhysRevLett.120.016602,turker2020bosonization}. 

The bosonization duality in relativistic field theories in 2+1 dimensions \cite{SEIBERG2016395,Karch_PhysRevX.6.031043} finds its precursors \cite{MPA-Fisher_PhysRevB.48.13749} in actual condensed matter systems where quasiparticles with fractional statistics (anyons) arise, like for example in the fractional quantum Hall effect \cite{ezawa2013quantum,Nayak_RevModPhys.80.1083}. Recall that in quantum Hall systems statistical transmutation occurs via an emergent gauge field associated to the Berry phase of the Laughlin wave function \cite{Arovas_PhysRevLett.53.722}. The idea is that bosons can be transmuted into fermions via flux attachment, which changes the (intrinsic) angular momentum from integer to half-integer via a mechanism reminiscent of the Aharonov-Bohm effect \cite{STERN2008204}. The emergent gauge field is typically not governed by an electrodynamics \`a la Maxwell, as in this case the flux alone cannot guarantee the change of statistics via the angular momentum, since a Maxwell-like electrodynamics produces a field contribution to the angular momentum that generally cancels the mechanical contribution arising by letting the particle go around another particle with a flux attached to it \cite{jackiw1990topics,Forte_RevModPhys.64.193}. That is the reason why a topological action in the form of a CS term is required: the energy-momentum tensor being given by the functional derivative of the action with respect to the metric, a CS term does not contribute to it, since it is topological (i.e., it does not depend on the metric) \cite{jackiw1990topics}, so no field contribution to the total angular momentum arises from it. 

Things become considerably more subtle when gauge fields are periodic, i.e., the $U(1)$ group is compact. In this case even the pure Abelian CS action,
\begin{equation}
	S_{CS} = \frac{i \kappa}{2}\int d^3x a_\mu \epsilon_{\mu \nu \lambda} \partial_{\nu} a_\lambda,
\end{equation} 
features nontrivial aspects, despite the theory being free. This situation provides one of the simplest realizations of a so-called topological order \cite{wen2004quantum}. The compactness of the gauge field $a_\mu$ forces the CS coupling to be quantized, $\kappa = k/(2\pi)$,  where $k \in \mathbb{Z}$ is the so-called CS level. This way eigenstates of the Wilson loop have a degeneracy that depends on the genus $g$ of the manifold where the CS form is defined, being given by $k^g$. This fact is not only important in a purely abstract sense, but has far reaching consequences in condensed matter systems in 2+1 dimensions \cite{wen2004quantum,fradkin2013field}. There are other, mathematically deeper ways of understanding the quantization of the CS action, and this involves concepts like spin manifolds and connections \cite{Seiberg-Witten-2016}. 

The Wilson loops and the topological quantization, which  they ensue in the pure compact CS action, acquire an even more interesting meaning when the loops are actually closed vortex lines due to the coupling to matter fields. Recall that in the (bosonic) particle-vortex duality \cite{Peskin1978,THOMAS1978513,Dasgupta-Halperin_PhysRevLett.47.1556,kleinert1982disorder} particle worldlines in a theory are expressed in terms of an ensemble of vortex lines in a dual theory, with the duality transformation mapping the strong coupling of a theory into the weak coupling of the other one, and vice-versa. In this case the topologically nontrivial gauge transformation involves the phase $\theta$ of the complex Higgs scalar, so $a_\mu\to a_\mu+\partial_\mu\theta$ causes the CS action to be transformed into $S_{CS}'[a]=S_{CS}[a+d\theta]$,
%
but periodicity of $\theta$ implies that $\epsilon_{\mu \nu \lambda}\partial_\nu\partial_\lambda\theta$ does not vanish at vortex singularities \cite{kleinert1989gauge}. Considering a single vortex loop for simplicity,  
\begin{equation}
	\epsilon_{\mu \nu \lambda}\partial_\nu\partial_\lambda\theta(x)=2\pi n\oint_L dy_\mu\delta^3(x-y),
\end{equation}
where $n\in\mathbb{Z}$ is the winding number defined via $\oint_L d\theta=2\pi n$, we obtain from the CS action a field equation for the gauge field in the form, 
\begin{equation}
	\epsilon_{\mu \nu \lambda}\partial_{\nu}a_\lambda=-2\pi kn\oint_L dy_\mu\delta^3(x-y).
\end{equation}       
The quantity $J_\mu=i\epsilon_{\mu \nu \lambda}\partial_{\nu}a_\lambda/(2\pi)$ defines a topological current. The role of topology is apparent also in the sense of the vortex loop as a topological defect, since the RHS turns out to have a zero divergence independently of the shape of the loop. 

In Ref. \cite{Peskin1978} the correlation functions of vortex loops are shown to be directly related to the current correlation function associated to the global (Noether) $U(1)$ symmetry. There this result was established on the lattice in the context of a bosonic particle-vortex duality. Such a connection is expected to exist also in the case of other dualities in 2+1 dimensions, including the ones associated to bosonization \cite{Goldman2018}. Particularly relevant in this context is the approach to the critical point, where the theory is a conformal field theory (CFT) and, in addition, the fixed points on both sides of the duality coincide. This example highlights well the power of the duality approach, since it establishes the existence of a critical point (and therefore a second-order phase transition) in the Abelian Higgs model in 2+1 dimensions for one Higgs scalar, something that is otherwise very hard to assert by perturbative means, unless a large enough number of scalars is available \cite{HLM_PhysRevLett.32.292,Hikami,Herbut_PhysRevB.100.134507}.   
The duality result obtained in Ref. \cite{Peskin1978} in the lattice and elaborated further via Monte Carlo simulations in Refs. \cite{Dasgupta-Halperin_PhysRevLett.47.1556,Hove-Sudbo_PhysRevLett.84.3426} allows to unequivocally conclude that a second-order phase transition indeed takes place. That this result can also be obtained within a field theoretical approach via bosonization is a remarkable achievement \cite{SEIBERG2016395,Karch_PhysRevX.6.031043}. 

Regarding the phase structure, the bosonization duality approach is considerably less conclusive in the case of a $U(1)\times U(1)$-symmetric Higgs theory in 2+1 dimensions  \cite{Senthil_PhysRevX.7.031051}. In this case numerical results using lattice gauge theory models favor a first-order phase transition \cite{KUKLOV20061602,Kragset_PhysRevLett.97.247201}. However, other lattice models with  $U(1)\times U(1)$ symmetry lead to a second-order phase transition \cite{Hove-Sudbo_PhysRevB.75.104405,Sudbo_PhysRevB.87.134503,Pollmann_PhysRevLett.120.115702}. In a recent letter \cite{Shyta_PhysRevLett.127.045701} we have pointed out yet another route towards quantum criticality in $U(1)\times U(1)$ Higgs systems, namely, one where a CS term is present in the Lagrangian. There we focused on a derivation of the duality in terms of interactions between vortex loops, which in the case of a CS Higgs theory naturally leads to a sum over sign factors $(-1)^{N_{ab}}$ in the partition function, where $N_{ab}=\sum_{a\neq b}n_an_b Lk_{ab}$, with integers $n_a$ and $n_b$ being the quanta (winding numbers) from given vortex loops $a$ and $b$. These sign factors are weighted by a phase factor $e^{i\pi\sum_an_a\mathcal{W}_a}$, where $\mathcal{W}_a$ is the writhe \cite{calugareanu1959integrale,calugareanu1961classes,Moffatt_doi:10.1098/rspa.1992.0159}. The sign factors are here identified to integrated out massless fermions. The duality and a renormalization group (RG) analysis provided strong evidence for the existence of a quantum critical point in this theory. Key in the RG analysis was the vanishing of certain Feynman diagrams at zero momenta as a consequence of the CS term \cite{Shyta_PhysRevLett.127.045701}. 

In this paper we discuss the $U(1)\times U(1)$ CS Higgs theory further using a point of view more focused on local field theory actions rather than directly considering vortex loop configurations. The basic idea is to take advantage of the flux attachment with CS terms in the  bosonic particle-vortex duality for the $U(1)$ theory \cite{Karch_PhysRevX.6.031043}. In spite of the absence of a CS term in the Higgs theory of this bosonic duality, a CS $U(1)$ Higgs model is used in the intermediate steps via the bosonization technique in 2+1 dimensions \cite{SEIBERG2016395,Karch_PhysRevX.6.031043}. Here we will discuss some important differences between the bosonization duality for the usual $U(1)\times U(1)$ Higgs theory discussed earlier \cite{Karch_PhysRevX.6.031043,Senthil_PhysRevX.7.031051} and the case of a CS $U(1)\times U(1)$ Higgs theory. One main focus of our analysis is to obtain results that elucidate further the quantum critical behavior discussed in our previous work \cite{Shyta_PhysRevLett.127.045701}. We will address with considerable level of detail the correspondence via the duality between the conserved current associated to the $U(1)\times U(1)$ symmetry and the topological current from the dual model. This point of view adheres to the logic of bosonization in (1+1)-dimensional systems \cite{Coleman_PhysRevD.11.2088} where a precise correspondence is derived between the currents of the fermionic and bosonic theories. For this reason we will start by briefly reviewing some basic facts about the (1+1)-dimensional case in Section \ref{Sec:Bos-1+1}.

Besides the duality and critical behavior of the $U(1)\times U(1)$ CS theory,  
this paper will additionally explore several aspects of the particle-vortex duality in 2+1 dimensions that reflect the interplay between current correlation functions featuring both Noether and topological currents. As mentioned, we will mainly focus on theories at the quantum critical point. In order to set the stage, after briefly recapping in Section \ref{Sec:Bos-1+1} well known results from bosonization in 1+1 dimensions, our discussion of bosonization in 2+1 dimensions in Section \ref{Sec:Bos-2+1} starts by first considering a toy example in 0+1 dimensions (Section \ref{Sec:Bos-0+1}). The interest in such an example is not only driven by the fact that it provides exact results within a path integral treatment. In fact, the (0+1)-dimensional case  exhibits many features similar to those encountered in the (2+1)-dimensional case, but in a much simpler setting. One of the similarities being the generation of a CS term in 0+1 dimensions after exactly performing the path integral over fermions \cite{Deser_PhysRevLett.79.1976,Dunne_PhysRevLett.78.3434}. Also the effect of the parity anomaly in the partition function and its role in the bosonization is made explicit. Given that the  Thirring model in 1+1 dimensions is one of the best known examples of bosonization \cite{Coleman_PhysRevD.11.2088}, we briefly consider it in both $0+1$ (Section \ref{Sec:Thirring-0+1}) and $2+1$ (Section \ref{Sec:Thirring-2+1}) dimensions. Sections \ref{Sec:CP-PV}, \ref{Sec:Flux}, \ref{Sec:BosDual2-comp}, and \ref{Sec:Disc} constitute the core results of this paper. Section \ref{Sec:CP-PV} focuses mainly on the role of criticality in particle-vortex duality, offering new insights from a field theoretic perspective on the well known bosonic particle-vortex duality in 2+1 dimensions \cite{Peskin1978,THOMAS1978513,Dasgupta-Halperin_PhysRevLett.47.1556}. There the importance of the critical point is emphasized in the discussion we give of this duality (Section \ref{Sec:PV-duality}). Next, establishing the bosonization duality for the $U(1)$-symmetric Higgs theory, in Section \ref{subsubsec:currcorrel} we obtain the relation between topological and Noether current correlators at criticality. Section \ref{Sec:Flux} serves as a warm-up for the  discussion of the $U(1)\times U(1)$ case in Sections \ref{Sec:BosDual2-comp} and \ref{Sec:Disc}. Section \ref{Sec:BosDual2-comp} gives a derivation of both the bosonic and fermionic dual models as well as the calculation of the corresponding current correlators. In Section \ref{Sec:Disc} we discuss the phase structure and compare it to the one of the $U(1)\times U(1)$ Higgs electrodynamics without the CS term. Section \ref{Sec:Conclusion} summarizes our results and gives an outlook.

\section{Bosonization in 1+1 dimensions}
\label{Sec:Bos-1+1}

Let us briefly recapitulate a number of results from (1+1)-dimensional bosonization that will be relevant within the context of this paper. 

\subsection{The Thirring model}

The Thirring model of interacting Dirac fermions represents a primary example of bosonization in field theory \cite{Coleman_PhysRevD.11.2088}. The Lagrangian for the Thirring model differs from the one for QED in the way the $U(1)$ Noether currents interact. Instead of featuring a long-range, photon-mediated interaction between currents, the interaction between currents in the Thirring model is short-ranged. The Lagrangian is thus given by, 
\begin{equation}
	\label{Eq:Thirring}
	\mathcal{L}_{\rm Th}=\bar{\psi}(\slashchar{\partial}+M)\psi-\frac{\lambda}{2}j_\mu j_\mu,
\end{equation} 
where the current, $j_\mu=i\bar{\psi}\gamma_\mu\psi$. In 1+1 dimensions the Thirring model is known to be dual to the bosonic sine-Gordon model, 
\begin{equation}
	\label{Eq:SG-L}
	\mathcal{L}_{\rm SG}=\frac{1}{2\kappa^2}(\partial_\mu\varphi)^2-\alpha\cos\varphi,
\end{equation} 
where $\alpha\sim M$. 
The conserved current $j_\mu$ of the Thirring Lagrangian is related to the scalar field $\varphi$ in the sine-Gordon Lagrangian via \cite{Coleman_PhysRevD.11.2088},
\begin{equation}
	\label{Eq:Th-curr-bos}
	j_\mu=\frac{1}{2\pi}\epsilon_{\mu\nu}\partial_\nu\varphi,
\end{equation} 
provided the following relation between couplings holds, 
\begin{equation}
	\label{Eq:kappa}
	\kappa^2=\frac{4\pi^2}{\pi+\lambda}. 
\end{equation}
Such a connection between the currents reflects the dual nature of the corresponding theories.  

In the zero mass limit the  Thirring model in 1+1 dimensions is equivalent to a free scalar field, since in this case $\alpha=0$. Thus, as elaborated in Appendix \ref{app:O(2)-nlsm}, we obtain that the current correlation function $C_{\mu\nu}(x)=\langle j_\mu(x)j_\nu(0)\rangle$ is given by, 
\begin{equation}
	\label{eq:thirr11Corr}
	C_{\mu\nu}(x)=-\frac{1}{2\pi(\pi+\lambda)x^2}\left(\delta_{\mu\nu}-\frac{2x_\mu x_\nu}{x^2}\right). 
\end{equation}
This result highlights the future importance of the  prefactor arising in the current correlator, which is dependent on the interaction coupling $\lambda$.  
\subsection{Fermions in the presence of a background gauge field}
\label{sec:1plus1Dirac}

There is yet another instance where a current having the form (\ref{Eq:Th-curr-bos}) plays a role. Let us consider the action of a massless Dirac fermion in the presence of a background gauge field, 
\begin{equation}
	\label{Eq:S-F}
	S=\int d^2 x\bar{\psi}(\slashchar{\partial}-i\slashchar{A})\psi.
\end{equation}
This action features besides the local $U(1)$ gauge symmetry also the {\it local} chiral symmetry $\psi(x)\to e^{i\gamma_5\varphi(x)}$, since this transformation is compensated by $A_\mu(x)\to A_\mu(x)-\epsilon_{\mu\nu}\partial_\nu\varphi$. This is true because in $d=2$ the property that $\gamma_\mu\gamma_5=\epsilon_{\mu\nu}\gamma_\nu$ holds.  Hence, the gauge field is entirely parameterized in terms of a scalar field. At the classical level a gauge transformation in the fermionic fields removes henceforth the gauge field completely from the action, leading to a theory of a free massless scalar fermion. As is well known, such a result is not correct at the quantum level, as the theory of Eq. (\ref{Eq:S-F}) features a chiral anomaly. Recall that the effective action, 
\begin{equation}
	\label{Eq:Seff}
	S_{\rm eff}=-\ln\det(\slashchar{\partial}-i\slashchar{A}),
\end{equation}
needs to be regularized. Using the standard Pauli-Villars regularization, one obtains that the scalar field parametrization of the gauge field cannot be gauged away any longer. The final result is a theory consisting of a free massless Dirac fermion decoupled from a free massless scalar field \cite{zinn2002quantum}, 
\begin{equation}
	\label{eq:1plus1PV}
	S=\int d^2x\left[\bar{\psi}\slashchar{\partial}\psi+\frac{1}{2\pi}(\partial_\mu\varphi)^2\right]. 
\end{equation} 

The scalar field term in the above action corresponds to the $O(2)$ nonlinear sigma model considered in detail in Appendix \ref{app:O(2)-nlsm} (with $g=\pi$). 
As the vector gauge field $A_\mu$ is eliminated, instead of the fermionic current $j_\mu = i \bar{\psi}\gamma_\mu\psi$ there is a corresponding topological current, $\tilde{J}_\mu=\frac{1}{\pi}\epsilon_{\mu\nu}\partial_\nu\varphi$, obtained from Eq. \eqref{eq:1plus1PV}. Analogously, the current $J_\mu=\partial_\mu\varphi/\pi$  is identified to the fermionic chiral current $j_\mu^5=i\bar{\psi}\gamma_5\gamma_\mu\psi$. Due to the chiral anomaly, the current correlation function for $J_\mu$ is not transverse in momentum space, as shown in Appendix \ref{app:O(2)-nlsm}. Therefore, through a proper Pauli-Villars regularization one obtains a bosonized version of the (1+1)-dimensional massless fermionic theory with a background gauge field.

\section{Dynamically generated Chern-Simons term and Bosonization}
\label{Sec:Bos-2+1}

Bosonization in 2+1 dimensions works very differently from the (1+1)-dimensional case. One distinctive feature is the absence of a $\gamma_5$ matrix and, as a consequence, a chiral symmetry in the case of massless fermions \footnote{Regarding this point, the following remark is in order. In some versions of QED in 2+1 dimensions a different version of the chiral symmetry can be introduced \cite{Appelquist_PhysRevD.33.3704}. In this case one considers four component Dirac fermions and a $\gamma_5$-like matrix, anti-commuting with all other Dirac $\gamma$ matrices can be defined. However, such a $\gamma_5$-like matrix is not given as a product of the other Dirac matrices. Such a version of QED in 2+1 dimensions has numerous applications in condensed matter physics, see for example Refs. \cite{Rantner-Wen_PhysRevLett.86.3871,Franz_PhysRevB.66.054535,Herbut_PhysRevLett.97.146401,Herbut_PhysRevB.66.094504,Kleinert-Nogueira_PhysRevB.66.012504,Kaul-Sachdev_PhysRevB.77.155105,Nogueira_PhysRevB.77.195101,Gusynin_PhysRevB.67.115201}. However, in this case there is no parity anomaly, since neither time-reversal nor spatial inversion are broken by adding a mass term. The latter actually leads to a breaking of the chiral symmetry (defined in the mentioned sense) instead}. Although in 2+1 dimensions the product of Dirac matrices does not yield a $\gamma_5$ matrix, the trace of such a product  produces ${\rm tr}(\gamma_\mu\gamma_\nu\gamma_\lambda)=2i\epsilon_{\mu \nu \lambda}$, which is of paramount importance in the dynamical generation of the CS term. The possibility of a CS term is another distinctive feature of systems at an odd number of spacetime dimensions. The simplest occurrence of a CS term is in 0+1 dimensions, where most calculations can be performed exactly. This also provides a toy model to illustrate some key features of the bosonization duality in 2+1 dimensions. Concerning the latter, an early example amounts to extend to 2+1 dimensions the (1+1)-dimensional bosonization of the Thirring model \cite{FRADKIN1994253}. We briefly review this result in Sect. \ref{Sec:Thirring-2+1} below, since some elements of it will be useful latter on. Also, for completeness, since the Thirring model is being discussed in both 1+1 and 2+1 dimensions, we connect the exact bosonization duality in 0+1 dimensions to a (0+1)-dimensional version of the Thirring model, which just turns out to correspond to the Hubbard model in the atomic limit.  

\subsection{Exact bosonization in 0+1 dimensions}
\label{Sec:Bos-0+1}

\subsubsection{Bosonization duality and parity anomaly in 0+1 dimensions}

The bosonization duality takes different forms depending on the dimensionality of the theories. In pursuit of a clear narrative behind the (2+1)$d$ bosonization that we are set to obtain, let us start by demonstrating an exact bosonization duality in (0+1)$d$ with only the time component present. In the imaginary time formalism, a fermionic theory  has the following Lagrangian, 
\begin{equation}
	\label{Eq:L-F-0+1}
	\mathcal{L}_f=\psi^\dagger(\partial_\tau+m-iA(\tau))\psi, 
\end{equation} 
where $\psi(\tau)$ is a fermionic field coupled to  the gauge field $A$ and satisfying an anti-periodic boundary condition, $\psi(0)=-\psi(\beta)$, with $\beta= 1/T$ being an inverse temperature.   While the Lagrangian is gauge invariant, the partition function fails to have this property when topologically nontrivial gauge transformations are considered. By the latter we mean a gauge transformation $\psi\to e^{i\lambda}\psi$, $A\to A+\partial_\tau\lambda$, where $\lambda\in S^1$. We will see how this works within this simple setting below.  

Once  the fermionic field is integrated out, the partition yields, 
\begin{equation}
	Z_f=\frac{\det(\partial_\tau+m-iA(\tau))}{\det(\partial_\tau+m)},
\end{equation}
where for normalization purposes there is a division by the free theory. The fermion determinant is given by the product of the eigenvalues of the operator $(\partial_\tau+m+A(\tau))$. The boundary conditions on the fermionic field allow to obtain a discretized set of eigenvalues, and thus, the partition function becomes, 
\begin{eqnarray}
	\label{Eq:Z}
	Z_f&=&\prod_{n=-\infty}^{\infty}\left[1+\frac{\beta m-i\phi}{i(2n+1)\pi}\right]\\&\times&\prod_{n=-\infty}^{\infty}\left[1+\frac{\beta m}{i(2n+1)\pi}\right]^{-1},\nonumber
\end{eqnarray}
where we used the notation,
\begin{equation}
	\label{Eq:0-d-flux}
	\phi=\int_{0}^{\beta}d\tau A(\tau).
\end{equation}

The infinite products arising in the partition function (\ref{Eq:Z}) can be evaluated by manipulating the well known infinite product representation of the cosine hyperbolic function \cite{arfken2005mathematical}. Eventually, one obtains following Refs.  \cite{Deser_PhysRevLett.79.1976,Dunne_PhysRevLett.78.3434}, 
\begin{equation}
	Z_f=\frac{\cosh\left(\frac{\beta m-i\phi}{2}\right)}{\cosh\left(\frac{\beta m}{2}\right)}=\cos\left(\frac{\phi}{2}\right)-i\tanh\left(\frac{\beta m}{2}\right)\sin\left(\frac{\phi}{2}\right).
\end{equation}
Assuming the mass, $m$, to be positive and taking the limit $\beta m\to \infty$, we obtain the zero temperature result, 
\begin{equation}
	\label{Eq:Zf-0+1}
	Z_f=e^{-\frac{i\phi}{2}}.
\end{equation}
Note that the zero temperature limit resembles a Pauli-Villars type of regularization, if the limit is interpreted as a large mass one. Importantly, if the ``flux" of Eq. (\ref{Eq:0-d-flux}) is $2\pi$, in analogy with the single unit flux in 3+1 dimensions, we obtain that $Z_f=-1$, so the partition function is negative. This result is a toy example for a more general scenario provided by the Atiyah-Patodi-Singer (APS) theorem \cite{atiyah_patodi_singer_1975}; see also Ref. \cite{Witten_RevModPhys.88.035001} for a discussion in the context of partition functions of fermionic systems in manifolds with and without boundaries.  

A topologically nontrivial gauge transformation, $A\to A +\partial_\tau\lambda$ yields, 
\begin{equation}
	Z_f'=(-1)^nZ_f,
\end{equation}
where $n$ is the winding number, 
\begin{equation}
	\label{Eq:Winding-n}
	n=\frac{1}{2\pi}\int_{0}^\beta d\tau\partial_\tau\lambda, ~~~~~n\in\mathbb{Z}. 
\end{equation} 

The above sign flip in the partition function that is caused by the gauge transformation is the (0+1)-dimensional version of the well known parity anomaly in 2+1 dimensions \cite{Redlich_PhysRevLett.52.18,Redlich_PhysRevD.29.2366}. By means of this toy example we can see that the anomaly is canceled at zero temperature if we add a level -1/2 CS term  to the action, which in (0+1)$d$ is simply a gauge field itself with an imaginary prefactor. The resulting theory without the anomaly is then given by, 
\begin{equation}
	\label{Eq:anomcancel}
	Z_f'(A)=\int\mathcal{D}\psi^\dagger\mathcal{D}\psi e^{-S_f+\frac{i}{2}\int_0^\beta d\tau A(\tau)}=1.
\end{equation}
A similar argument like the one above plays an important role in (2+1)$d$, as we will see a couple of times later on. 

Although we have considered a massive theory, we note that the anomaly persists in the massless case, since for $m=0$ we have, 
\begin{equation}
	Z_f|_{m=0}=\cos(\phi/2),
\end{equation}
which again gets a $(-1)^n$ factor after performing a topologically nontrivial gauge transformation. 


Now we consider the bosonic Lagrangian in (0+1) dimensions containing a level 1 CS term, 
\begin{equation}
	\mathcal{L}_b=\frac{1}{2g}(\partial_\tau\theta-A)^2+iA,
\end{equation}
where $\theta\in S^1$. This Lagrangian can be regarded as a topological Abelian Higgs model in 0+1 dimensions. 

Performing the shift $A\to A +\partial_\tau\theta$ we obtain, 
\begin{equation}
	\label{Eq:L-shift-0+1}
	\mathcal{L}_b'=\frac{1}{2g}A^2+i(A+\partial_\tau\theta),
\end{equation}
and thus in view of Eq. (\ref{Eq:0-d-flux}), 
\begin{equation}
	Z_b=e^{-i\phi},
\end{equation}
after using a proper normalization. The important result for us is the phase factor.  
From the discussion for the fermion case, we can therefore write, 
\begin{eqnarray}
	&\int\mathcal{D}\psi^\dagger\mathcal{D}\psi e^{-S_f-\frac{i}{2}\int_0^\beta d\tau A(\tau)}
	\nonumber\\
	&=\int\mathcal{D}\theta e^{-\int_0^\beta d\tau\left[\frac{1}{2g}(\partial_\tau\theta-A)^2+iA\right]}.
\end{eqnarray}
The above is the simplest example of the bosonization duality involving CS terms. We note in particular a feature that will appear later in our discussion of bosonization in 2+1 dimensions, namely, that the fermionic side features a level 1/2 CS term, while a level 1 CS occurs in the bosonic side.  

\subsubsection{The Thirring model in 0+1 dimensions}
\label{Sec:Thirring-0+1}

Given the results of the previous subsection, we see that if we promote $A$ to be a dynamical gauge field $a$ over which we integrate over in the path integral, the Lagrangian of Eq. (\ref{Eq:L-F-0+1}) actually imposes a constraint $\psi^\dagger\psi=0$. Thus, inspired also by Eq. (\ref{Eq:L-shift-0+1}), we add a term $a^2/(2g)$ to the Lagrangian of Eq. (\ref{Eq:L-F-0+1}). The result obtained upon integrating out $a$ is an interacting fermionic theory, 
\begin{equation}
	\label{Eq:Hubbard-atomic}
	\mathcal{L}'_f=\psi^\dagger(\partial_\tau+m)\psi+\frac{g}{2}(\psi^\dagger\psi)^2,
\end{equation}
where in the present case we must require that there are at least two Fermi field components, otherwise $(\psi^\dagger\psi)^2$ vanishes, since $\psi$ is a Grassmann field. The Lagrangian (\ref{Eq:Hubbard-atomic}) corresponds to the Thirring model in 0+1 dimensions. 
In the two-component case [for instance, $\psi=(\psi_\uparrow,\psi_\downarrow)$], the Lagrangian (\ref{Eq:Hubbard-atomic}) is equivalent to the atomic limit of the Hubbard model. In this two-component example where $a$ arises as an auxiliary field (via a so called Hubbard-Stratonovich transformation), integrating out the fermions yields in the zero temperature limit, 
\begin{equation}
	Z=\int\mathcal{D}ae^{-i\phi-\frac{1}{2g}\int d\tau a^2(\tau)}.  
\end{equation}
The partition function above presents a bosonized version of the Thirring model. 

To build up the solid groundwork for the discussion of the (2+1)-dimensional case, we also need general current correlation functions of massless scalar and fermionic theories. The expressions are known, but for the sake of completeness of the narrative these results are derived in Appendix \ref{app:ints}. With this and bosonization in lower dimensions in mind, we transition to the  bosonization occurring in   (2+1)$d$. As before, the conversation is opened by the fermionic case. While there are similarities when approaching the (1+1)$d$ and (2+1)$d$ cases, as was hinted above they are not fully analogues. For instance, the same action of Eq. (\ref{Eq:S-F}) taken in 2+1 dimensions does not have a chiral symmetry, since $\gamma_5$ matrices can only be defined for even dimensions of spacetime. However, just as in 0+1 dimensions, the theory  features a parity anomaly, since the effective action (\ref{Eq:Seff}) in 2+1 dimensions receives a factor $(-1)^n$ under a topologically nontrivial gauge transformation, with $n$ being the winding number introduced in Eq. \eqref{Eq:Winding-n} and associated to the nontrivial topology \cite{Redlich_PhysRevD.29.2366}.

\subsection{The Thirring model in 2+1 dimensions}
\label{Sec:Thirring-2+1}

A bosonization procedure for the Thirring model in 2+1 dimensions consists in performing a 
straightforward  Hubbard-Stratonovich transformation to introduce 
an auxiliary vector field $a_\mu$, 
\begin{equation}
	{\cal L}'_{\rm Th}=\bar{\psi}(\slashchar{\partial}+i\slashchar{a}+M)\psi+\frac{1}{2\lambda}a_\mu^2.
\end{equation} 
Integrating out fermions generates the 
following effective Lagrangian \cite{FRADKIN1994253} in the regime where the mass $M$ is large, 
\begin{equation}
	\label{Eq:Leff}
	{\cal L}_{\rm eff}=\frac{1}{2\lambda}a_\mu^2+\frac{i{\rm sgn}(M)}{8\pi}\epsilon_{\mu\nu\lambda}a_\mu\partial_\nu a_\lambda+\frac{1}{48\pi|M|}(\epsilon_{\mu \nu \lambda}\partial_\nu a_\lambda)^2.
\end{equation} 
This effective Lagrangian follows directly from the large mass expansion of the current correlation function for free fermions of Eq. (\ref{Eq:CC-F}) in Appendix \ref{app:free-fermions}, since it also corresponds to the one-loop vacuum polarization.

A couple of remarks are in order. First, we note that the Lagrangian (\ref{Eq:Leff}) is not gauge invariant because of the term $\sim a_\mu^2$. However, even in the absence of such a term (for instance, in the limit $\lambda\to\infty$), gauge invariance is lost if one allows for topologically nontrivial gauge transformations, since the CS term is half-quantized in the above calculation \cite{Redlich_PhysRevLett.52.18}. 
Strictly speaking, such large gauge transformations are not required in the case of the Thirring model, since $a_\mu$ here is a non-compact auxiliary field. Nevertheless, let us take the opportunity to consider this strong coupling regime and the possibility of allowing for large gauge transformations, since these are of paramount importance in most contexts considered in this paper.  
In fact, this leads us to consider Pauli-Villars regularization in the form, 
\begin{equation}
	\label{Eq:PauliVillars}
	S_{\text{eff}}=-\operatorname{Tr} \ln (\slashchar{\partial}+M-i\slashchar{a})+  \operatorname{Tr} \ln (\slashchar{\partial}-M-i\slashchar{a}),
\end{equation}
assuming $M>0$ infinitely large such as to suppress the Maxwell term and turn the generated CS term integer-quantized. In this way, we obtain, 
\begin{equation}
	\label{Eq:Leff-1}
	{\cal L}'_{\rm eff}=\frac{1}{2\lambda}a_\mu^2+\frac{i}{4\pi}\epsilon_{\mu\nu\lambda}a_\mu\partial_\nu a_\lambda,
\end{equation}
which is invariant under topologically nontrivial gauge transformations when $\lambda\to\infty$.     

The field equation for the effective Lagrangian (\ref{Eq:Leff-1}) expresses that the theory is self-dual (in the sense of field equations) \cite{TOWNSEND198438,DESER1984371},
\begin{equation}
	a_\mu=-i\frac{\lambda}{2\pi}\epsilon_{\mu \nu \lambda}\partial_\nu a_\lambda.
\end{equation} 
The above self-duality may be explored further by introducing an auxiliary field $b_\mu$ to rewrite (\ref{Eq:Leff-1}) as, 
\begin{equation}
	\mathcal{L}''_{\rm eff}=\frac{\lambda}{8\pi^2}(\epsilon_{\mu \nu \lambda}\partial_\nu b_\lambda)^2+\frac{i}{4\pi}\epsilon_{\mu \nu \lambda}a_\mu\partial_\nu\left( a_\lambda-b_\lambda\right).
\end{equation}
Integrating out $a_\mu$ yields then the equivalent dual effective Lagrangian, 
\begin{equation}
	\widetilde{L}_{\rm eff}=\frac{\lambda}{8\pi^2}(\epsilon_{\mu \nu \lambda}\partial_\nu b_\lambda)^2-\frac{i}{4\pi}\epsilon_{\mu \nu \lambda}b_\mu\partial_\nu b_\lambda,
\end{equation} 
which is gauge invariant. Hence, similarly to the (1+1)-dimensional case, the massive Thirring model offers a path to bosonization in 2+1 dimensions as well. However, the result is not exact in the latter case. Nevertheless, we see already here that the level 1 CS term appears naturally in the (2+1)-dimensional bosonization to preserve the gauge invariance.  This subsection highlights the interplay between gauge noninvariance, parity anomaly and the role CS term plays in both of them.

\section{The critical point and particle-vortex duality}
\label{Sec:CP-PV}
By this point, we touched on the correspondence between the current correlation functions occurring in (1+1)-dimensional dualities in Section \ref{Sec:Bos-1+1}. We now generalize this discussion as well as examine the (2+1)-dimensional case in detail. Current correlators describe the behavior of a theory in the near-critical regime, and so as dual theories are expected to behave similarly the notions of criticality and duality are intertwined. This section will demonstrate this connection through the bosonic particle-vortex duality as well as consider some of the more intricate procedures involved.
\subsection{Scale invariance and current conservation}

It is well known that scale invariance and current conservation imply that the scaling behavior of the current correlation function does not have an anomalous dimension \cite{collins1985renormalization,Kovtun_PhysRevD.78.066009}. This means that if a theory features a critical point, the power-law behavior of the current correlation function is determined by dimensional analysis. Thus, the power law obtained for the free theory is also valid for the interacting case at an RG fixed point, therefore, only the amplitude is changed. To account for this possibility, we generalize the current correlation functions of a free massless bosonic theory in momentum space calculated in Eq. \eqref{Eq:C-1-p} of Appendix \ref{app:ints} to be, 
\begin{equation}
	\label{Eq:CC-general}
	C_{\mu\nu}(q)=-v c_d|q|^{d-2}\left(\delta_{\mu\nu}-\frac{q_\mu q_\nu}{q^2}\right),
\end{equation}
where $v$ is a universal amplitude equal to unity for a free scalar theory and $c_d$ is a coefficient defined by the dimensionality of the theory and given explicitly in Eq. \eqref{Eq:cd}.

Interestingly, as is shown in Appendix \ref{app:ints}, the case of $v=1$ also corresponds to the one-loop vacuum polarization of the critical Higgs theory that we will encounter further in Eq. (\ref{Eq:SQED}). This one-loop value changes for a multicomponent Higgs theory. For instance, in the case of a $O(2N)$-invariant theory, where $N$ is the number of complex fields, we have, $v=N$, since one has to sum over all $N$ complex scalar loops. This is the well-known leading order large $N$ result.  

As was discussed in the previous section, fermionic currents are intrinsically dependent on the structure of $\gamma_\mu$ matrices, therefore, the generalization to $d$ dimensions when calculating the current correlation functions is tiresome. Nevertheless, in 2+1 dimensions the current correlation function of free massless Dirac fermions coincides with the bosonic one presented in Eq. \eqref{Eq:CC-general} for $v=1$ (see Appendix \ref{app:ints}). The crucial feature of the $d=3$ case lies in the fact that in the most general situation one  has to account for a possibility of the parity symmetry breaking in the form of a CS term \cite{Witten-2003},
%
\begin{equation}
	\label{Eq:CC-general-p}
	C_{\mu\nu}(p)=-\frac{|p|}{16}\left[v\left(\delta_{\mu \nu}-\frac{p_\mu p_\nu}{p^2}\right)+w\epsilon_{\mu \nu \lambda}\frac{p_\lambda}{|p|}\right],
\end{equation}    
where $w$ is another universal amplitude.  This equation will be shown to be consistent with our original calculations in 2+1 dimensions in Sections \ref{Sec:Flux} and \ref{Sec:BosDual2-comp}. 

\subsection{Particle-vortex duality and the existence of a critical point}
\label{Sec:PV-duality}

Current correlation functions at criticality were shown to yield universal results reflecting the duality as fermionic and bosonic correlators possess the same form. Let us demonstrated now how duality can be used as a tool to claim the existence of a critical point.  Particle-vortex duality in 2+1 dimensions is performed exactly on the lattice using a Villain form of the $XY$ model \cite{Peskin1978,THOMAS1978513}. In the version discussed originally by Peskin \cite{Peskin1978}, it is shown that the $XY$ model is equivalent by duality to a lattice Abelian Higgs model with infinite phase stiffness in Euclidean spacetime. In a classical statistical mechanics sense this was treated as a zero temperature limit in Ref. \cite{Peskin1978}, which was dubbed "frozen superconductor". On the other hand, the temperature of the original $XY$ model becomes the inverse of the gauge coupling constant $e^2$ in the dual model. This line of reasoning ultimately led to the conclusion in a later work by Dasgupta and Halperin \cite{Dasgupta-Halperin_PhysRevLett.47.1556} that the Abelian Higgs model (in the form of a lattice Ginzburg-Landau superconductor \cite{kleinert1989gauge}) features a critical point, thus undergoing a second-order phase transition. This result is very hard to obtain using RG methods (see, for instance, Ref. \cite{Herbut_PhysRevB.100.134507} for a recent account and the closing paragraph of this subsection). More sophisticated numerical studies \cite{Hove-Sudbo_PhysRevLett.84.3426} have confirmed the duality scenario of earlier papers \cite{Peskin1978,THOMAS1978513,Dasgupta-Halperin_PhysRevLett.47.1556,kleinert1982disorder}, including an analysis of the scaling dimensions. This convincingly confirms by means of lattice simulations the validity of the scale invariant form of the current correlation function as given in Eq. (\ref{Eq:CC-general-p}) with $w=0$ (i.e., corresponding to the situation where parity symmetry breaking is absent). Furthermore, since the renormalization of the gauge coupling is well known to be given in terms of the vacuum polarization by, 
\begin{equation}
	e^2(p)=\frac{e^2}{1+e^2\Pi(p)},
\end{equation}
Eq. (\ref{Eq:CC-general-p}) implies, $e^2(p)=e^2/[1+e^2v/(16|p|)]$. We obtain that the dimensionless coupling $\hat{e}^2(p)=e^2(p)/|p|$ reaches the IR stable fixed point, $\hat{e}^2_*=16/v$, as $|p|\to 0$. We conclude that this theory is strongly coupled as $e^2\to\infty$ causes the dimensionless coupling to flow to the same fixed point. In this case, the bare coupling constant $e^2$ serves as a UV scale. The same line of argument holds for fermionic QED in 2+1 dimensions \cite{Appelquist_PhysRevD.33.3704}.

The discussion above allows us to establish the field theory version of the bosonic particle-vortex duality on the lattice as the equality of the partition functions in the IR regime, i.e., the one in which $e^2\to\infty$, so the Maxwell term can be suppressed \footnote{However, the Maxwell term is frequently needed for obtaining more rigorous results, since it provides a UV regulator, also within the bosonization duality context involving Chern-Simons actions. See, for instance, Ref. \cite{Shyta_PhysRevLett.127.045701}}. In order to account for the current response, a background field has to be included in the corresponding partition functions. The way to proceed is already thoroughly discussed in Peskin's seminal work \cite{Peskin1978} and leads to the inclusion of what is now known as a BF term,
\begin{equation}
	\label{Eq:BF-term}
	S_{BF}[a;A]=\frac{i}{2\pi}\int d^3 xa_\mu\epsilon_{\mu\nu\lambda}\partial_\nu A_\lambda, 
\end{equation}
where $A_\mu$ denotes a background gauge field and $a_\mu$ represents a dynamical gauge field. As the field theory action associated to the $XY$ model is simply a $|\phi|^4$ field theory \cite{zinn2002quantum}, this bosonic field theory in the presence of background field $A_\mu$ is defined via,
\begin{eqnarray}
	\label{Eq:ZbQED}
	Z_{bQED}[A]&=&\int\mathcal{D}\phi^*\mathcal{D}\phi e^{-S_{bQED}[A]},
	\\
	S_{bQED}[A]&=&\int d^3x\left[|(\partial_\mu-iA_\mu)\phi|^2+m^2|\phi|^2+\frac{\lambda}{2}|\phi|^4
	\right].\nonumber
\end{eqnarray}
When the background field vanishes, $A_\mu=0$, the partition function $Z_{\rm bQED}$ is just the field theory version of the $XY$ model, and we can write $Z_{bQED}[A=0]=Z_{XY}$. Following the prescription above regarding an additional BF term, we have \cite{SEIBERG2016395,Karch_PhysRevX.6.031043},
\begin{eqnarray}
	\label{Eq:FT-Peskin}
	&&Z_{bQED}[A]=\int\mathcal{D}\phi^*\mathcal{D}\phi e^{-S_{bQED}[A]}
	\nonumber\\
	&=&\lim_{e^2\to\infty}\int\mathcal{D}\varphi^*\mathcal{D}\varphi\mathcal{D}a_\mu e^{-\int d^3x\mathcal{L}_{\rm sQED}[a]+S_{BF}[a;A]}
	\nonumber\\
	&=&Z_{sQED}[A]
\end{eqnarray} 
where 
\begin{equation}
	\label{Eq:SQED}
	\mathcal{L}_{\rm sQED}[a]=\frac{1}{4e^2}F_{\mu\nu}^2+|(\partial_\mu-ia_\mu)\varphi|^2+m^2|\varphi|^2+\frac{u}{2}|\varphi|^4
\end{equation}
is the usual scalar QED Lagrangian. Note the difference in the two QED models yielding partition functions denoted $Z_{bQED}[A]$ and $Z_{sQED}[A]$. In the former the fluctuating scalar field is coupled minimally to a (classical) background field $A_\mu$, while in the latter {\it both} the scalar and the minimally coupled gauge field are being integrated over. In $Z_{sQED}[A]$ the background field couples to the fluctuating gauge field via a BF term.  

Equation (\ref{Eq:FT-Peskin}) is interesting, since it establishes the particle-vortex duality as something that mimics dual variables in ordinary quantum mechanics, the most prosaic example being given by the duality between the position and momentum operators via the commutator, $[x,p]=i$, which is also reflected in terms of a plane wave basis via Fourier transforms of wave functions. In the functional integral of Eq. (\ref{Eq:FT-Peskin}) the phase factor contains the BF term, which plays a role similar to the phase factor in an ordinary Fourier transform underlying the particle-wave duality in quantum mechanics. In this sense, the partition function of Eq. (\ref{Eq:ZbQED}) acquires the Fourier representation in terms of fluxes as represented in Eq. (\ref{Eq:FT-Peskin}). Note that by promoting $A_\mu$ to a dynamical gauge field $b_\mu$ on the RHS of Eq. (\ref{Eq:FT-Peskin}), adding a BF term coupling $b_\mu$ to a new background field $B_\mu$, trivially yields the LHS upon subsequently integrating out $b_\mu$, which enforces $a_\mu=B_\mu$ (note that this simple calculation can also be done assuming $e^2$ finite). 

Along a similar vein, a fermionic analog of the bosonic particle-vortex duality expressed in Eq. (\ref{Eq:FT-Peskin}) has been proposed \cite{Son_PhysRevX.5.031027,Metlitski_PhysRevB.93.245151}. In the absence of parity symmetry breaking this amounts to a replacement of the scalar QED by a fermionic QED in Eq. (\ref{Eq:FT-Peskin}), along with the switching to corresponding path integrals over spinor (Grassmann) fields. 

From Eq. (\ref{Eq:FT-Peskin}) it follows that the current correlation function of the $XY$ model field theory,
\begin{equation}
	C_{\mu\nu}(x-x')=\left.\frac{\delta\ln Z_{bQED}[A]}{\delta A_\mu(x)\delta A_\nu(x')}\right|_{A=0},
\end{equation}
is connected to the correlation function $\mathcal{K}_{\mu\nu}(x-x')=\langle J_\mu(x)J_\nu(x')\rangle$ of the topological current,
\begin{equation}
	\label{Eq:Top-curr}
	J_{\mu}=\frac{i}{2 \pi} \varepsilon_{\mu \nu\lambda} \partial_{\nu} a_{\lambda}.
\end{equation}
These correlation functions will be explored in a more quantitative fashion in the next section. 

Let us discuss the limit $e^2\to\infty$ further. Integrating out $a_\mu$ exactly on the RHS of Eq. (\ref{Eq:FT-Peskin}) yields the effective Lagrangian, 
\begin{eqnarray}
	\label{Eq:Leff-no-a}
	\mathcal{L}_{\rm eff}&=&|\partial_\mu\varphi|^2+m^2|\varphi|^2+\frac{u}{2}|\varphi|^4
	\nonumber\\
	&+&\frac{\left(\varphi^*\partial_\mu\varphi-\varphi\partial_\mu\varphi^*-\frac{\epsilon_{\mu \nu \lambda}\partial_\nu A_\lambda}{2\pi}\right)^2}{4|\varphi|^2}.
\end{eqnarray}
When the background field is set to zero, the last term of Eq. (\ref{Eq:Leff-no-a}) represents a (perturbatively) non-renormalizable interaction in 2+1 dimensions, as near a $U(1)$ broken symmetry state the denominator $|\varphi|^2\sim{\rm const}$ at leading order. This is an irrelevant operator in the RG sense, so essentially upon integrating $a_\mu$ out in the RHS of Eq. (\ref{Eq:FT-Peskin}) we indeed obtain (asymptotically) its LHS.  

The discussion of the previous paragraph glosses over the effect of the functional measure arising in the path integral after integrating out $a_\mu$ in the limit $e^2\to\infty$. In fact, after integrating out $a_\mu$ the measure becomes formally, $\mathcal{D}\varphi^*\mathcal{D}\varphi/|\varphi|$ (implicit in this expression is a formal infinite product of the form, $\prod_x [d\varphi^*(x)d\varphi(x)/|\varphi(x)|]$).  Although this has no major consequences perturbatively, it will be important in the following manipulations of the path integral. 

Let us now express the complex scalar field in terms of an amplitude and a phase as, $\varphi=\rho e^{i\theta}/\sqrt{2}$, with the result, 
\begin{eqnarray}
	\label{Eq:Leff-rho-theta}
	\mathcal{L}_{\rm eff}&=&\frac{1}{2}\left[(\partial_\mu\rho)^2+m^2\rho^2+\frac{u}{4}\rho^4\right]
	\nonumber\\
	&+&\frac{1}{8\pi^2\rho^2}(\epsilon_{\mu \nu \lambda}\partial_{\nu} A_\lambda)^2-\frac{i}{2\pi}\partial_\mu\theta\epsilon_{\mu \nu \lambda}\partial_{\nu}A_\lambda,
\end{eqnarray}
while the path integral measure becomes $\mathcal{D}\varphi^*\mathcal{D}\varphi/|\varphi|=\prod_x[d\rho(x)d\theta(x)]$. 
 Partial integration of the last term of the effective Lagrangian (\ref{Eq:Leff-rho-theta}) in the action leads to a direct coupling of $A_\mu$ to loops of vortex lines, since periodicity of $\theta$ implies that the spacetime curl of $\partial_\mu\theta$ does not vanish along vortex lines, yielding instead \cite{kleinert1989gauge}, 
\begin{equation}
	\epsilon_{\mu \nu \lambda}\partial_{\nu}\partial_\lambda\theta=2\pi\sum_an_a\oint_{L_a} dy_\mu^{(a)}\delta^3(x-y^{(a)}),
\end{equation}
where $n_a\in\mathbb{Z}$ and $L_a$ denote vortex loop paths. Thus, the Wilson loop factors arise in the partition function,  
\begin{equation}
	\prod_a e^{i2\pi n_a\oint_{L_a}dy_\mu^{(a)}A_\mu(y^{(a)})}, 
\end{equation}
with the functional integral over the phase $\theta(x)$ becoming a path integral over all possible particle trajectories along the loops $L_a$. Therefore, the current correlation function can be expressed in terms of a vortex loop correlation function, i.e., a correlation function between topological defects. The lattice version of this statement can already be encountered in Ref. \cite{Peskin1978}. Here we are rephrasing it in terms of a continuum field theory approach, which also highlights the role played by amplitude fluctuations with respect to the background field: from Eq. (\ref{Eq:Leff-rho-theta}) we note that $\rho^2$ plays the role of a gauge coupling in the Maxwell term for the background field.

The correctness of the effective Lagrangian (\ref{Eq:Leff-rho-theta}) can be cross checked by showing that it indeed yields the LHS of Eq. (\ref{Eq:FT-Peskin}) after  promoting the background field $A_\mu$ to a dynamical gauge field $b_\mu$, adding a BF term $S_{BF}[b;B]$ in the process, with $B_\mu$ being a new background field. We have already seen that this is a trivial calculation if done before integrating out $a_\mu$, since $b_\mu$ in this case simply acts as a Lagrange multiplier field. It should not matter whether this is done before or after integrating out $a_\mu$. To see this, we  integrate out $b_\mu$ exactly, but the presence of the coefficient $1/\rho^2(x)$ multiplying the Maxwell term makes it necessary to proceed with some care. First, we insert the identity, 
\begin{equation}
	\int\mathcal{D}s_\mu\delta(s_\mu-\epsilon_{\mu \nu \lambda}\partial_{\nu}b_\lambda)=1,
\end{equation}
and use the integral representation of the delta function above, which makes a Lagrange multiplier field $\lambda_\mu$ appear in the action,
\begin{eqnarray}
	\mathcal{L}_{\rm eff}&=&\frac{1}{2}\left[(\partial_\mu\rho)^2+m^2\rho^2+\frac{u}{4}\rho^4\right]
	\nonumber\\
	&+&\frac{1}{8\pi^2\rho^2}s_\mu^2+is_\mu\left(\frac{B_\mu}{2\pi}-\frac{\partial_\mu\theta}{2\pi}-\lambda_\mu\right)
	\nonumber\\
	&+&i\lambda_\mu\epsilon_{\mu \nu \lambda}\partial_\nu b_\lambda.
\end{eqnarray} 
Integrating out $b_\mu$ leads to the constraint $\epsilon_{\mu \nu \lambda}\partial_{\nu}\lambda_\mu=0$, which is immediately solved as $\lambda_\mu=\partial_\mu\omega$ and trivially gauged away via $\theta\to\theta-2\pi\omega$. Finally, integrating over $s_\mu$  
yields, 
\begin{equation}
	\mathcal{L}_{XY}=\frac{1}{2}\left[(\partial_\mu\rho)^2+m^2\rho^2+\frac{u}{4}\rho^4\right]+\frac{\rho^2}{2}\left(\partial_\mu\theta-B_\mu\right)^2,
\end{equation}
while restoring the Jacobian factor back to $\prod_x[d\rho(x) \rho(x)d\theta(x)]$. The result can obviously be rewritten back in the form (\ref{Eq:ZbQED}) by reintroducing the complex fields via $\phi=\rho e^{i\theta}/\sqrt{2}$. 

 We note here the power of the duality in establishing the existence of a critical point, something otherwise difficult to obtain by purely perturbative means \cite{Folk_PhysRevB.41.4083,Folk_1996,Herbut_PhysRevB.100.134507} (for a non-perturbative analysis of the problem using the functional RG, see Refs. \cite{Wetterich-IJMP-A,Wetterich_PhysRevB.53.5734}). Although  the existence of an IR fixed point for a properly defined (i.e.,in terms of some chosen scaling variable) dimensionless gauge coupling is easy to obtain at one-loop order, the same does not hold with the $|\phi|^4$ coupling, since the one-loop calculation requires a number $N$ of complex field components to be larger than 183 \cite{HLM_PhysRevLett.32.292,Hikami}. This critical number of fields gets significantly reduced in a four-loop calculation, but an IR stable fixed point is still not found even in this case \cite{Herbut_PhysRevB.100.134507}. Extensive numerical simulations were made in a recent work \cite{Bonati_PhysRevB.103.085104} for several values of $N$ (specifically, $N=2, 4, 10, 15, 25$), starting with $N=2$, conclusively finding a second-order phase transition for $N\geq 10$. Importantly, a first-order phase transition is found for the $N=2$ case \cite{Bonati_PhysRevB.103.085104}, leading to an agreement with earlier RG results \cite{HLM_PhysRevLett.32.292,Hikami}. Thus, it is quite remarkable that the field theory duality can establish the existence of a critical point for the single complex field case in the strong-coupling regime $e^2\to\infty$. As a side remark, let us point out that perturbatively even the RG flow for the dimensionless gauge coupling introduces difficulties at higher loop order. For instance, while at one-loop a an IR fixed point is found for any number of Higgs scalars, the same fails to be true at two-loops: the $\epsilon$-expansion for the dimensionless gauge coupling yields a fixed point provided more than 18 Higgs fields are available \cite{Folk_PhysRevB.41.4083,Folk_1996}. However, the duality result implies that such an IR fixed point must exist non-perturbatively for a single Higgs scalar. This in turn implies the scaling behavior of the current correlation function. Most importantly, no detailed information is needed about the RG flow of the $|\phi|^4$ coupling, which underlies most of the difficulties involving the establishment of Wilson-Fisher critical behavior \cite{HLM_PhysRevLett.32.292,Hikami,Herbut-Tesanovic_PhysRevLett.76.4588,Herbut_PhysRevB.100.134507}.  Importantly, the vanishing of the RG $\beta$ function for the gauge coupling is well known to exactly determine the anomalous dimension of the gauge field, which is in this case given by $4-d$ \cite{Herbut-Tesanovic_PhysRevLett.76.4588,Wetterich_PhysRevB.53.5734,deCalan-Nogueira_PhysRevB.60.4255,Hove-Sudbo_PhysRevLett.84.3426}. Thus, since the gauge field couples linearly to the conserved current, the absence of an anomalous dimension in the latter follows immediately, leading in this way to a current correlator given by Eq. (\ref{Eq:CC-general}).   

\subsection{From background to dynamical field via fluctuating vortex loops}
\label{subsec:vortex-loop-A}   
The discussion of the particle-vortex duality in  the previous subsection involved the   promotion of the background gauge field to a dynamical one. Let us explore the meaning of this procedure  and   physical motivation behind it in terms of vortex loops.
 
In a regime where $\rho$ is approximately uniform, we can adopt a different point of view about promoting $A_\mu$ to a dynamical field that emphasizes the role of vortex loops. In this sense, promoting $A_\mu$ to a dynamical gauge field corresponds to solving the field equation for $A_\mu$ as, 
\begin{equation}
	\label{Eq:A-background}
	A_\mu(x)=B_\mu(x)+i\frac{\rho^2}{2}\sum_a n_a\oint_{L_a}\frac{dy_\mu^{(a)}}{|x-y^{(a)}|},
\end{equation}
where $B_\mu(x)$ is a non-singular classical background contribution that does not depended on vortex loops. 
At first sight, the fact that this solution is complex might seem problematic, since $A_\mu$ is supposed to be real. However, this is actually not a problem at all, since the imaginary contribution is interpreted in the sense of quantum fluctuations in an ensemble of vortex loops. After all, performing a Gaussian integral in this case amounts to simply solving the field equations and plugging the solution (\ref{Eq:A-background}) back into the action. In this way the Wilson loop term generates a contribution that yields the interaction between vortex loops in the effective action, 
\begin{eqnarray}
	\label{Eq:Vortex-loops-int}
	&S_{\rm Vortex-loop-int}=\pi\rho^2\sum_{a,b}n_an_b
	\nonumber\\
	&\times\oint_{L_a}dy_\mu^{(a)}\oint_{L_b}dy_\mu^{(b)}\frac{1}{|y^{(a)}-y^{(b)}|}.
\end{eqnarray} 
Such a vortex interaction term naturally appears in lattice Villain models \cite{Savit_RevModPhys.52.453}. In the particular case of Eq. (\ref{Eq:Vortex-loops-int}), it just corresponds to the continuum version of the vortex loop representation of the Villain form of the $XY$ model in 2+1 dimensions  \cite{Dasgupta-Halperin_PhysRevLett.47.1556}.  
The analysis above emphasizes at the end the precise meaning of the particle-vortex duality, since we ultimately have to perform a path integral over vortex loops that are direct worldlines of particle trajectories. Also, this subsection, in particular Eq. \eqref{Eq:A-background}, motivates the notion of promoting a background gauge field to be dynamical. 

%
%
 
%

\section{Bosonization duality and critical current correlators in the $U(1)$-symmetric case}
\label{Sec:Flux}
In the previous section we demonstrated how the existence of a critical regime can follow from the particle-vortex duality.  As the end-goal of this work is to examine the bosonization duality between the CS $U(1)\times U(1)$ Higgs theory and interacting massless fermions, this section sets up the necessary methods on a simpler model. Here we will obtain the bosonization duality for the CS $U(1)$-symmetric Higgs theory employing the flux attachment approach. Moving further, we will show the way in which the duality is reflected in the relation between current correlation functions corresponding to different sides of the duality. 
\subsection{Bosonization via flux attachment}
\label{Subsubsec:Flux}

%
The flux attachment technique is very powerful in its simplicity as it uses   the two basic bosonization equations as building blocks of the duality web \cite{Karch_PhysRevX.6.031043,Raghu_PhysRevLett.120.016602},
\begin{equation}
	\label{Eq:Boson+flux}
	Z_{f Q E D}[A] e^{-\frac{1}{2}S_{CS}[A]}=Z_{b Q E D+flux}[A],
\end{equation}
\begin{equation}
	\label{Eq:Fermion+flux}
	Z_{f Q E D+flux}[A]=Z_{b Q E D}[A] e^{S_{CS}[A]},
\end{equation}
where $S_{CS}[A]$ is the action for a level 1 CS term,
\begin{equation}
	\label{eq:CSterm}
	S_{CS} = \frac{i}{4\pi} \epsilon_{\mu\nu\lambda}A_\mu\partial_\nu A_\lambda,
\end{equation}
and $A_\mu$ is the background field. Let us describe the elements arising in Eqs. (\ref{Eq:Boson+flux}) and (\ref{Eq:Fermion+flux}) in detail. While the bosonic partition function was introduced in Eq. \eqref{Eq:ZbQED}, the fermionic side is given by, 
\begin{eqnarray}
	\label{Eq:ZfQED}
	Z_{fQED}[A]&=&\int\mathcal{D}\bar{\psi}\mathcal{D}\psi e^{-S_{fQED}[A]},
	\nonumber\\
	S_{fQED}[A]&=&\int d^3x\bar{\psi}(\slashchar{\partial}-i\slashchar{A})\psi.
\end{eqnarray}
The fermionic flux attachment is defined as,    
\begin{eqnarray}
	\label{Eq:ZfQED+flux}
	Z_{fQED+flux}[A]&=&\int\mathcal{D}a_\mu 	Z_{fQED}[a]e^{-\frac{1}{2}S_{CS}[a] -S_{BF}[a;A]},
	\nonumber\\
\end{eqnarray}
while for bosons one has,
\begin{eqnarray}
	\label{Eq:ZbQED+flux}
	Z_{bQED+flux}[A]&=&\int\mathcal{D}a_\mu Z_{bQED}[a]e^{S_{CS}[a] + S_{BF}[a;A]},
\nonumber\\
\end{eqnarray}
with the BF term defined in Eq. \eqref{Eq:BF-term}.

Equation (\ref{Eq:Fermion+flux}) can easily be shown to follow from Eq. (\ref{Eq:Boson+flux}) \cite{tong2018gauge}, so actually the latter can be regarded as the fundamental result out of which all other dualities belonging to the duality web are derived \cite{SEIBERG2016395}. 
%
Conversely, a similar reasoning would allow us to derive Eq. (\ref{Eq:Boson+flux}) starting from (\ref{Eq:Fermion+flux}), so it would have been equally possible to assume the latter as a seed to all other dualities as well. This way one can obtain the exact particle-vortex duality  discussed in Sect. \ref{Sec:PV-duality} from  one of the flux attachment conjectures and its time-reversed version. Such an elegant approach is also consistent with other well established dualities and can be used to provide new reliable results. 

Note that the flux attachment for fermions operates slightly differently from the bosonic case. Indeed, while the bosons get a level 1 CS term, the fermionic flux attachment features a level 1/2 CS term. The reason for this is the parity anomaly \cite{Redlich_PhysRevLett.52.18,Redlich_PhysRevD.29.2366}. We have encountered this problem before in our discussion of the Thirring model in 2+1 dimensions. It is now the moment of being more thorough on this point. It turns out that massless Dirac fermions are invariant under TR and inversion of one of the two spatial coordinates, but not both, since in two spatial dimensions this would correspond to a rotation by $\pi$. In Euclidean three-dimensional spacetime both symmetries translate into parity symmetry, corresponding to an inversion of any of the three coordinates. It turns out that quantum fluctuations (i.e., integrating out the fermions in a gauge field background) would either break gauge invariance or the parity \cite{Redlich_PhysRevD.29.2366}. The solution to this problem is well known. Just like in the (1+1)-dimensional case, we have to employ Pauli-Villars regularization in the fermionic theory. 

The effective action for free massless fermions can be obtained assuming dimensional regularization for the parity even component of current correlation function (as shown in Eq. (\ref{Eq:CC-F}) in Appendix \ref{app:ints}). Nevertheless, this type of regularization is not suitable to obtain a CS term in the massless case. In order to do so, 
let us consider the theory with an additional level 1/2 CS gauge term and integrate out fermionic fields. In this case, the following effective action is obtained
\begin{equation}
	\label{eq:ferm21PauliVillars}
	S_{\text{eff}}=-\operatorname{Tr} \ln (\slashchar{\partial}-i\slashchar{A})+  \operatorname{Tr} \ln (\slashchar{\partial}-M-i\slashchar{A})+\frac{1}{2} S_{C S}(A),
\end{equation}
where $M>0$. The second tracelog in Eq. (\ref{eq:ferm21PauliVillars}) arises from the Pauli-Villars regulator fields. When the limit $M\to\infty$ is taken the regularized theory yields an additional level 1/2 CS contribution that results in an overall level 1 CS term. Were we add a level -1/2 CS term, no CS term at all would be present in the effective action. We have noticed a similar behavior within the simple setting in 0+1 dimensions, see, for instance, Eq. (\ref{Eq:anomcancel}). Note that nothing like this happens in the bosonic theory, since scalar field fluctuations do not generate any anti-symmetric contribution in the vacuum polarization.  

Curiously, from Eqs. (\ref{Eq:FT-Peskin}) and (\ref{Eq:Fermion+flux}) we can write, 
\begin{eqnarray}
	\label{Eq:FB-duality}
	&&Z_{f Q E D+flux}[A] e^{-S_{CS}[A]}=Z_{b Q E D}[A]
	\nonumber\\
	&=&\int\mathcal{D}a_\mu Z_{b Q E D}[a]e^{S_{BF}[a;A]},
\end{eqnarray} 
so we note that for $A_\mu=0$ there must be no parity symmetry breaking occurring in $Z_{f Q E D+flux}[A=0]$, in spite of the (dynamical) CS term in Eq. (\ref{Eq:ZfQED+flux}), since $Z_{b Q E D}[A=0]$ does not break parity symmetry. This is consistent with the discussion above. Although the classical fermionic action seem to break parity, the latter is actually not broken at the quantum level \cite{SEIBERG2016395}.  

\subsection{Current correlation functions}
\label{subsubsec:currcorrel}

With the duality \eqref{Eq:Boson+flux} in mind, let us calculate the current correlators for bosonic and fermionic sides. We will start with the former by first integrating over the scalar field $\phi$ at one-loop order in Eq. (\ref{Eq:ZbQED+flux}). Then, the CS $U(1)$ Higgs action can be rewritten as a quadratic form,
\begin{equation}
	\label{eq:21bosflux}
	S_B = \frac{1}{2}\int_p a_\mu (p) M_{\mu \nu}(p) a_\nu(-p),
\end{equation}
with
\begin{equation}
	M_{\mu \nu}(p)=\frac{|p|}{16}\left(\delta_{\mu\nu}-\frac{p_{\mu} p_{\nu}}{p^{2}}\right)-\frac{1}{2 \pi} \varepsilon_{\mu\nu \lambda} p_{\lambda}+\frac{|p|}{\alpha} \frac{p_{\mu} p_{\nu}}{p^{2}},
\end{equation}
where the last term contains a gauge fixing constant $\alpha$. Inverting the tensor we obtain a propagator for $a_\mu$,
\begin{equation}
	\label{Eq:Dmunu}
	D_{\mu \nu} (p) = \frac{\pi^2}{4\left(\frac{\pi^2}{64} + 1\right)|p|} \left(\delta_{\mu \nu} - \frac{p_\mu p_\nu}{p^2} + \frac{8}{\pi |p|} \varepsilon_{\mu \nu \lambda} p_\lambda\right),
\end{equation}
written here in the Landau gauge $\alpha=0$. 

The theory features the topological current of Eq. (\ref{Eq:Top-curr}), and 
one can thus see that the current correlation function in this case is directly derived from the propagator $D_{\mu \nu}$ and yields in momentum space,
\begin{eqnarray}
	\label{Eq:CC-B-massless}
	\mathcal{K}_{\mu\nu}^B(p)=\left\langle J_{\mu}(p) J_{\nu}(-p)\right\rangle &=& -\frac{1}{\frac{\pi^{2}}{64}+1}\left[\frac{|p|}{16}\left(\delta_{\mu \nu}-\frac{p_{\mu} p_{\nu}}{p^{2}}\right)\right.
	\nonumber\\
	&+&\left.\frac{1}{2 \pi} \varepsilon_{\mu \nu \lambda} p_{\lambda}\right].
	\label{eq:21bosonCS}
\end{eqnarray}
%

The obtained topological current correlator has to be compared to the one on the fermionic side of the bosonization duality in Eq. \eqref{Eq:Boson+flux}. To do so, we recall that the current correlation function of a free massless fermion is given by Eq. \eqref{Eq:CC-general-p} with $v=1$ and $w=0$ (see Eq. \eqref{Eq:CC-F-massless} in Appendix \ref{app:free-fermions}). Note that an anti-symmetric contribution to the current correlator is generated by the proper Paulli-Villars regularization as shown in Eq. (\ref{eq:ferm21PauliVillars}). Finally, the fermionic side of the duality at hand contains an additional level 1/2 CS term. Therefore, the fermionic current correlation function reads, 
\begin{equation}
	\label{eq:masslessferm2plus1}
	\mathcal{C}^F_{\mu\nu}(p)=-\frac{|p|}{16}\left(\delta_{\mu \nu}-\frac{p_{\mu} p_{\nu}}{p^{2}}\right)-\frac{1}{2\pi}\epsilon_{\mu \nu \lambda}p_\lambda.
\end{equation}
Comparing the above expression to the general scale invariant form in Eq. (\ref{Eq:CC-general-p}), we see that $v=1$ and $w=8/\pi$. 
Hence, the topological current correlation function is proportional to $\mathcal{C}^F_{\mu\nu}(p)$, 
\begin{equation}
	\label{eq:2plus1CCFrelation}
	\mathcal{K}_{\mu\nu}^B(p)=\frac{1}{\frac{\pi^2}{64}+1}\mathcal{C}^F_{\mu\nu}(p).
\end{equation}
As was conveyed in Section \ref{Sec:CP-PV}, such a relation between the current correlators not only supports the validity of the bosonization duality but also points towards the existence of a critical regime for both theories in the duality. 

The interpretation of the coefficient relating topological and Noether current correlators is a well known manifestation of  S-duality in 2+1 dimensions \cite{Witten-2003}: the coefficients of the transverse and parity odd components of $\mathcal{K}_{\mu\nu}^B(p)$ are the real and imaginary parts of $1/(4\pi^2\tau)$, where $\tau=-1/16+i/(2\pi)$ is viewed as a complex-valued coupling constant that gets inverted by the duality. The real and imaginary parts of $\tau^*$ are the transverse and parity odd coefficients of $\mathcal{C}^F_{\mu\nu}(p)$. 
 
 \section{Bosonization duality in $U(1)\times U(1)$ topological Abelian Higgs model}
 \label{Sec:BosDual2-comp}
 We have seen by now that bosonization duality for the one-component CS Higgs model results in current correlation functions on both sides of the relation coinciding up to a universal prefactor. We showed in the previous section how the duality between bosonic and fermionic theories in 2+1 dimensions can be obtained via flux attachment. Now, the discussion above will finally result in the derivation of the bosonization duality involving the $U(1)\times U(1)$ topological Abelian Higgs model. We will also calculate corresponding current correlation functions and conclude whether their form implies the existence of a critical point in the duality. To elaborate on the motivation to study such a duality, let us first put this model in the condensed matter physics context.
 
Some quantum antiferromagnets in two spatial dimensions are describable by means of a CP$^1$ model with an additional Maxwell term \cite{Motrunich-Vishwanath_PhysRevB.70.075104,Senthil1490,Sachdev-book}. In this case the direction of the magnetization $\vec{n}$ is given in the terms of the CP$^1$ fields $z_a$ ($a=1,2$) by $\vec{n}=z_a^*\sigmab_{ab}z_b$. Such a model is known to be self-dual when an additional easy-plane anisotropy term, $Kn_z^2/2=K(|z_1|^2-|z_2|^2)^2/2$, $K>0$ is included in the Lagrangian, thus making the theory $U(1)\times U(1)$-symmetric \cite{Motrunich-Vishwanath_PhysRevB.70.075104}. Originally derived in the lattice \cite{Motrunich-Vishwanath_PhysRevB.70.075104}, this self-duality was rederived in Ref.  \cite{Karch_PhysRevX.6.031043} by means of the bosonization duality technique described in Sec. \ref{Sec:Flux}. 

Recently we have considered a version of such a quantum antiferromagnet with an additional CS term \cite{Shyta_PhysRevLett.127.045701}. One of the goals of that study was to show that the CS term causes the model to undergo a second-order phase transition, unlike the case where the CS term is absent \cite{KUKLOV20061602,Kragset_PhysRevLett.97.247201}. Specifically, the theory has a Lagrangian, 
\begin{eqnarray}
	\label{eq:CP1soft}
	\mathcal{L}&=&\sum_{a=1,2}|(\partial_\mu-ia_\mu)z_a|^2
	+\frac{i}{4\pi}\epsilon_{\mu\nu\lambda}a_\mu\partial_\nu a_\lambda
    \nonumber\\
	&+&m_0^2(|z_1|^2+|z_2|^2)+\frac{u}{2}(|z_1|^4+|z_2|^4),
\end{eqnarray}
where $a_\mu$ is an emergent gauge field. The Lagrangian above may also be viewed as a generalization of the theory of Eq. (\ref{Eq:ZbQED+flux}) to include two complex fields rather than a single one. This observation will be useful in the discussion that will follow later on. 

The theory is dual to another bosonic theory, which was established through the standard particle-vortex duality \cite{Shyta_PhysRevLett.127.045701}, which closely parallels the dualities performed in the lattice, though it was carried out directly in the continuum. The main reason for using a continuum approach is related to the technical difficulties one faces while trying to write a CS term in the lattice \cite{ELIEZER1992118,BERRUTO2000366,demarco2019lattice}. 
The dual bosonic field theory obtained in Ref. \cite{Shyta_PhysRevLett.127.045701} has the following form  yielding a second-quantized representation for the ensemble of vortex loops \cite{kleinert1989gauge},
 \begin{eqnarray}
 	\label{Eq:Dual-L}
 	\mathcal{L}_{dual}&=&\sum_{I=1,2}\left[|(\partial_\mu-ib_{I\mu})\phi_I|^2+m^2|\phi_I|^2+\frac{\lambda}{2}|\phi_I|^4
 	\right]\nonumber\\
 	&-&\frac{i}{4\pi}\epsilon_{\mu\nu\lambda}(b_{1\mu}+b_{2\mu})\partial_\nu (b_{1\lambda}+ b_{2\lambda}).
 \end{eqnarray}

We can establish the bosonization duality leading to Eq. (\ref{Eq:Dual-L}) using the flux attachment technique discussed in Sec. \ref{Sec:Flux}. We use the basic flux attachment dualities (\ref{Eq:Boson+flux}) and (\ref{Eq:Fermion+flux}) to derive the bosonization duality for the Lagrangian of Eq. (\ref{eq:CP1soft}) by simply multiplying both these relations together, 
 \begin{eqnarray}
 	\label{eq:zmult}
 	&Z&_{f Q E D}[A] e^{-\frac{1}{2}S_{CS}[A]} 	Z_{f Q E D+flux}[A]\nonumber\\=&Z&_{b Q E D+flux}[A]Z_{b Q E D}[A] e^{S_{CS}[A]},
 \end{eqnarray}
 After we promote the background field $A_\mu$ to a dynamical field $b_\mu$, the left-hand side of the Eq. \eqref{eq:zmult} takes the form,
 \begin{eqnarray}
 	&&\int\mathcal{D}b_\mu Z_{f Q E D}[b] e^{-\frac{1}{2}S_{CS}[b]} 	Z_{f Q E D+flux}[b]\nonumber\\
 	&=&\int\mathcal{D}b_\mu Z_{b Q E D+flux}[b]Z_{b Q E D}[b] e^{S_{CS}[b]},
\end{eqnarray}
which upon renaming the gauge fields, $a_\mu\to b_{1\mu}$ [remember that there is an integral over $a_\mu$ in $	Z_{f Q E D+flux}[A]$, recall Eq. (\ref{Eq:ZfQED+flux})], $b_\mu\to b_{2\mu}$, yields the fermionic action, 
\begin{eqnarray}
	\label{Eq:fAction}
 	S_f&=&\int d^3x\left\{\sum_{I=1,2}\left[ \bar{\psi}_I(\slashchar{\partial}-i\slashchar{b}_I)\psi_I+\frac{i}{8\pi}b_{I\mu}\epsilon_{\mu\nu\lambda}\partial_\nu b_{I\lambda}\right] \right.\nonumber\\
 	&+& \left.\frac{i}{2\pi}b_{1\mu}\epsilon_{\mu\nu\lambda}\partial_\nu b_{2\lambda}\right\}.\nonumber\\
 \end{eqnarray}
Proceeding similarly for the right-hand side of the  Eq. \eqref{eq:zmult}, we obtain that a bosonic action with the Lagrangian (\ref{Eq:Dual-L}). 
Therefore, we have obtained that the  derivation given via flux attachments to fermions and bosons reproduces the result (\ref{Eq:Dual-L}) derived in Ref. \cite{Shyta_PhysRevLett.127.045701} using particle-vortex duality in terms of vortex loops.  

While promoting the background field to a dynamical gauge field we have intentionally not introduced a new background field. The reason for this is that what has been done above would not have given the current response of the original model of Eq. (\ref{eq:CP1soft}) directly, since both Eqs. (\ref{Eq:Dual-L}) and (\ref{Eq:fAction}) are models dual to (\ref{eq:CP1soft}). What we need is to derive the dual fermionic model (\ref{Eq:fAction}) directly from the original action (\ref{eq:CP1soft}). In this way introducing background fields in an appropriate fashion leads ultimately to a relationship between the conserved current response of the model (\ref{eq:CP1soft}) to the topological current response of the fermionic dual model (\ref{Eq:fAction}).    
Hence, we will now show that the fermionic dual theory follows by starting directly from the Lagrangian  \eqref{eq:CP1soft}. This is achieved by first considering a time-reversed version of the Eq. \eqref{Eq:Fermion+flux}, 
\begin{equation}
	\label{Eq:TRFermion+flux}
	\bar{Z}_{f Q E D+flux}[A]=Z_{b Q E D}[A] e^{-S_{CS}[A]},
\end{equation}
where time reversal changes the signs in the flux attachment as follows,
\begin{equation}
	\label{Eq:TRZfQED+flux}
	\bar{Z}_{fQED+flux}[A]=\int\mathcal{D}a_\mu 	Z_{fQED}[a]e^{\frac{1}{2}S_{CS}[a]+S_{BF}[a;A]}.
\end{equation}
Then, the expression in Eq. \eqref{Eq:TRFermion+flux} is used twice in the following way, 
\begin{eqnarray}
	&&\bar{Z}_{f Q E D+flux}[A-C]\bar{Z}_{f Q E D+flux}[A+C]e^{S_{CS}[A-C]}\nonumber\\&=&Z_{b Q E D}[A-C] Z_{b Q E D}[A+C]e^{-S_{CS}[A+C]}.
\end{eqnarray}
Both sides of the equation are then multiplied by a BF term introducing a new background gauge field $B$, 
\begin{widetext}
	\begin{eqnarray}
		\label{eq:duality4_with_a}
		&&\int\mathcal{D}a_{\mu}Z_{\rm bQED}[a-C]Z_{bQED}[a+C]e^{-S_{CS}[a+C] + S_{BF}[a;B]{\color{black}-S_{CS}[B]}} \\&=& \int\mathcal{D}a_{\mu}\mathcal{D}b_{1\mu}\mathcal{D}b_{2\mu}Z_{\rm fQED}[b_1]Z_{\rm fQED}[b_2]e^{\frac{1}{2}\left(S_{CS}[b_1]+S_{CS}[b_2]\right)+S_{BF}[b_1;a+C]+S_{BF}[b_2;a-C]+S_{CS}[a-C]+S_{BF}[a;B]{-\color{black}S_{CS}[B]}}, \nonumber
	\end{eqnarray}
	where the background field $A$ was promoted to be dynamical $A\to a$ {\color{black}and we have included a CS term for the background field $B$ on both sides of Eq. (\ref{eq:duality4_with_a})}. After $a$ is integrated out in the RHS leading to $a = C - (b_1+ b_2+B)$, we obtain,
	\begin{eqnarray}
		\label{eq:duality4_0}
		&&\int\mathcal{D}a_{\mu}Z_{\rm bQED}[a-C]Z_{bQED}[a+C]e^{-S_{CS}[a+C] + S_{BF}[a;B]{\color{black}-S_{CS}[B]}} \\&=& \int\mathcal{D}b_{1\mu}\mathcal{D}b_{2\mu}Z_{\rm fQED}[b_1]Z_{\rm fQED}[b_2]e^{-\frac{1}{2}\left(S_{CS}[b_1]+S_{CS}[b_2]\right)-S_{BF}[b_1;b_2]+2S_{BF}[b_1;C] -S_{BF}[b_1+b_2;B]-{\color{black}2}S_{CS}[B]+S_{BF}[B;C]}. \nonumber
	\end{eqnarray}
Setting $B=C$, the duality takes the form,
\begin{eqnarray}
		\label{eq:duality4_1}
	Z_B[B]&\equiv&\int\mathcal{D}a_{\mu}Z_{\rm bQED}[a-B]Z_{bQED}[a+B]e^{-S_{CS}[a] - 2S_{CS}[B]} \\&=& \int\mathcal{D}b_{1\mu}\mathcal{D}b_{2\mu}\left(\prod_{I=1,2}\mathcal{D}\bar{\psi}_I\mathcal{D}\psi_I\right)e^{-S_f +S_{BF}[b_1-b_2;B]} \equiv Z_F[B]. \nonumber
\end{eqnarray}
\end{widetext}
The first line of Eq. (\ref{eq:duality4_1}) reproduces precisely the path integral for the Lagrangian of Eq. \eqref{eq:CP1soft} with bosons additionally coupled to the background field $B_\mu$. The second line features the fermionic action $S_f$ of Eq. (\ref{Eq:fAction}). Both sides of the expression contain additional terms coupling dynamical fields to the background field $B_\mu$. The notations $Z_B[B]$ and $Z_F[B]$ are introduced to refer to the bosonic and fermionic sides of the expression. It is clear, therefore, that Eq. (\ref{eq:duality4_1}) expresses an explicit bosonization duality for the theory in Eq. (\ref{eq:CP1soft}).  

Differentiating $\ln Z_B[B]=\ln Z_F[B]$ with respect to the background field, $B_\mu$, and setting $B_\mu=0$ at the end, one obtains the equality between the expectation values of the currents from both sides of the duality. The fermionic current is a topological one, 
\begin{equation}
	\label{Eq:jF}
	j^F_\mu=\frac{i}{2\pi}\epsilon_{\mu \nu \lambda}\partial_\nu (b_{2\lambda}-b_{1\lambda}),
\end{equation}
while the bosonic side yields the conserved current associated to the $U(1)\times U(1)$ symmetry, 
\begin{equation}
	\label{Eq:jB}
	j^B_\mu=\sum_{a=1,2}(-1)^a[i(z_a^*\partial_\mu z_a-z_a\partial_\mu z_a^*)+2|z_a|^2a_{\mu}].
\end{equation}

For current correlation functions for both bosonic and fermionic sides of the duality \eqref{eq:duality4_1} one can write, 
\begin{eqnarray}
	\label{Eq:Kf=Cb}
	\mathcal{C}^B_{\mu\nu}(x,x')&=&\left.\frac{\delta^2\ln Z_B[B]}{\delta B_\mu(x)\delta B_\nu(x')}
	\right|_{B=0},\nonumber\\
	\mathcal{K}^F_{\mu\nu}(x,x')&=&\left.\frac{\delta^2\ln Z_F[B]}{\delta B_\mu(x)\delta B_\nu(x')}\right|_{B=0}.
\end{eqnarray} 
It is anticipated that the current correlation functions are equal up to an overall universal amplitude prefactor. Note that for the fermionic correlation function non-diagonal gauge field propagators like $\langle b_{1\mu}(p)b_{2\nu}(-p)\rangle$ are needed. To obtain these we have to calculate the effective gauge action by integrating the fermions out. There is a simpler way, however. We can integrate the fermions out {\it before} we integrate $a_\mu$ out, i.e., using Eq. (\ref{eq:duality4_with_a}) instead. Following this procedure generate an additional $e^{\frac{1}{2}\left(S_{CS}[b_1]+S_{CS}[b_2]\right)}$ factor in the integrand. The integrals over $b_{1\mu}$ and $b_{2\mu}$ have the same (Gaussian) form, and thus we obtain upon integrating over both $b_{1\mu}$ and $b_{2\mu}$ the effective action, 
\begin{widetext}
\begin{eqnarray}
	\label{Eq:Seff-2fermions-with_a}
	S_{\rm eff}'=\frac{1}{4\pi^2}\int_x\int_{x'}\bar{D}_{\mu\nu}(x-x')\epsilon_{\mu\alpha\beta}\epsilon_{\nu\sigma\rho}\partial_\alpha\partial'_\sigma\left[a_\beta(x)a_\rho(x')+B_\beta(x)B_\rho(x')\right]-S_{CS}[a],
\end{eqnarray}
\end{widetext}
where in momentum space $\bar{D}_{\mu\nu}(p)=D_{\mu\nu}(-p)$, with $D_{\mu\nu}(p)$ being given by Eq. (\ref{Eq:Dmunu}), and we have set $B=C$ as before. By applying the prescription given in Eq. (\ref{Eq:Kf=Cb}), we obtain, 
\begin{equation}
	\mathcal{K}_{\mu\nu}^F(x-x')=-\left.\frac{\delta^2S_{\rm eff}'}{\delta B_\mu(x)\delta B_\nu(x')}\right|_{B=0},
\end{equation}
and therefore,
\begin{equation}
	\label{eq:finalFerm21}
	\mathcal{K}_{\mu\nu}^F(p)=-\frac{8|p|}{\pi^2+64}\left[\delta_{\mu\nu}-\frac{p_\mu p_\nu}{p^2}-\frac{8}{\pi}\frac{\epsilon_{\mu \nu \lambda}p_\lambda}{|p|}\right].
\end{equation}
Similarly, upon integrating out the scalar fields $z_a$ (at one-loop), Eq. (\ref{Eq:Kf=Cb}) yields, 
\begin{equation}
	\mathcal{C}_{\mu\nu}^B(p)=-\frac{|p|}{8}\left[\delta_{\mu\nu}-\frac{p_\mu p_\nu}{p^2}-\frac{8}{\pi}\frac{\epsilon_{\mu \nu \lambda}p_\lambda}{|p|}\right].
\end{equation}
Henceforth we verify the relation,
\begin{equation}
	\label{Eq:Kf=Cb-1}
	\mathcal{K}^F_{\mu\nu}(p)=\frac{1}{\frac{\pi^2}{64}+1}\mathcal{C}_{\mu\nu}^B(p).
\end{equation}

The universal amplitude relating $\mathcal{C}_{\mu\nu}^B(p)$ to $\mathcal{K}^F_{\mu\nu}(p)$ in Eq. (\ref{Eq:Kf=Cb-1}) is the same as the one appearing in Eq. (\ref{eq:2plus1CCFrelation}), with the difference that there the topological current correlator refers to the bosonic side of the duality and the $U(1)$ conserved current correlator is associated to the fermionic side of the duality. Actually what determines the nature of the current correlators (i.e., either a conserved current associated to global symmetry or a topological current) is how each side of the duality is coupled to the background field. In Eq. (\ref{Eq:FB-duality}) the background field $A$ is coupled to a $U(1)$ current on the fermionic side of the duality while it couples to a topological current on the bosonic side. As a consequence, Eq. (\ref{eq:2plus1CCFrelation}) ultimately follows from this fact. In the duality of Eq. (\ref{eq:duality4_1}), on the other hand, the background field is coupled to the $U(1)\times U(1)$ conserved current in the bosonic side of the duality, while the fermionic side couples the background field to the topological current. Another interesting aspect in this comparison is the presence of a CS term for the background field on the side where the latter is coupled to the conserved current following from a global (Noether) symmetry. In this case the coefficient of the CS term reflects the effective $U(1)$ charge of the theory. We note that for the Lagrangian (\ref{eq:CP1soft}) we can define an operator $z_1^*z_2$ that creates a monopole excitation of charge $+1-(-1)=2$, which reflects the level 2 of the CS term in Eq. (\ref{eq:duality4_1}).      

Working directly with the fermionic side of Eq. (\ref{eq:duality4_1}) where $a_\mu$ has already been integrated out is more cumbersome, since we have to deal with a matrix gauge field propagator with matrix elements, $\langle b_{I\mu}(p)b_{J\nu}(-p)\rangle$. While this is not strictly necessary for the calculation of the current correlators, it is important for obtaining other dynamical physical properties (we will encounter a situation in the next section where this happens, namely, in the study of the fermion self-energy as obtained from the Schwinger-Dyson equation). Let us therefore discuss this matrix propagator below. 

From the one-loop fermionic vacuum polarization, Eq. (\ref{eq:duality4_1}) leads to the effective gauge field action in the form,  
\begin{widetext}
\begin{eqnarray}
	\label{eq:matrixM}
	S_{\rm eff}^{\rm gauge}&=&\frac{1}{2}\int_p\left[
	\begin{array}{cc}
		b_{1\mu}(p) & b_{2\mu}(p)
	\end{array}
\right]
\left[
\begin{array}{cc}
	\frac{|p|}{16}\left(\delta_{\mu \nu}-\frac{p_\mu p_\nu}{p^2}\right) & -\frac{1}{2\pi}\epsilon_{\mu \nu \lambda}p_\lambda\\
	\noalign{\medskip}
	-\frac{1}{2\pi}\epsilon_{\mu \nu \lambda}p_\lambda & \frac{|p|}{16}\left(\delta_{\mu \nu}-\frac{p_\mu p_\nu}{p^2}\right)
\end{array}
\right]
\left[
\begin{array}{c}
	b_{1\nu}(-p)\\
	\noalign{\medskip}
	b_{2\nu}(-p)
\end{array}
\right]
\nonumber\\
&+&\frac{1}{2\pi}\int_p\epsilon_{\mu \nu \lambda}p_\lambda [b_{2\mu}(p)-b_{1\mu}(p)]B_\nu(-p),
\end{eqnarray}
\end{widetext}
\textcolor{black}{where the CS terms cancel due to the  level 1/2 CS terms with opposite signs generated by  integrating out fermions. }

From Eq. \eqref{eq:matrixM} one derives the matrix gauge field propagator in the Landau gauge, 
\begin{equation}
	\label{Eq:Matrix-gauge}
	\mathbb{D}_{\mu\nu}(p)=\left[
	\begin{array}{cc}
		D_{\mu\nu}(p) & E_{\mu\nu}(p)\\
		\noalign{\medskip}
		E_{\mu\nu}(p) & D_{\mu\nu}(p)
	\end{array}
	\right],
\end{equation}
where diagonal elements of the matrix are, 
\begin{eqnarray}
	\label{Eq:D-b1b1}
	&&D_{\mu\nu}(p)=\langle b_{1\mu}(p)b_{1\nu}(-p)\rangle=\langle b_{2\mu}(p)b_{2\nu}(-p)\rangle
	\nonumber\\
	&=&
	\frac{16 \pi^2}{|p|(\pi^2 + 64)}\left(\delta_{\mu \nu}-\frac{p_\mu p_\nu}{p^2}\right),
\end{eqnarray} 
and the off-diagonal elements can be calculated to be,
\begin{equation}
	E_{\mu\nu}(p)=\langle b_{1\mu}(p)b_{2\nu}(-p)\rangle
	=\frac{128 \pi}{(\pi^2 + 64)p^2}\epsilon_{\mu \nu \lambda}p_\lambda.
\end{equation}
The topological fermionic current correlation function corresponding to Eq. \eqref{Eq:jF} written in terms of the matrix propagator elements is,
\begin{equation}
	\mathcal{K}_{\mu\nu}^F(p)=-\frac{1}{2\pi^2}\epsilon_{\mu\alpha\beta}\epsilon_{\nu\sigma\rho}p_\alpha p_\sigma \left[D_{\beta\rho}(p) - E_{\beta\rho}(p)\right].
\end{equation} 
After a lengthy but straightforward calculation, one obtains once more 
the result 
of Eq. \eqref{eq:finalFerm21}.

\section{Discussion: bosonization and criticality in $U(1)\times U(1)$ symmetric theories}
\label{Sec:Disc}

As the $U(1)\times U(1)$ Abelian Higgs model that is most often considered in the literature does not contain a CS term in its Lagrangian \cite{Motrunich-Vishwanath_PhysRevB.70.075104,Senthil1490,Senthil_PhysRevB.70.144407}, both its bosonic \cite{Motrunich-Vishwanath_PhysRevB.70.075104} and fermionized \cite{Karch_PhysRevX.6.031043} forms of particle-vortex duality behave very differently relative to the model discussed in the previous section. Thus,  we will focus the discussion on the critical behavior of the topological two-component Higgs model and how it compares to the non-topological version.  First, from the standard Villain lattice form of particle-vortex duality \cite{Motrunich-Vishwanath_PhysRevB.70.075104} we easily infer that the effective continuum dual theory has the form \footnote{We have dropped the Maxwell terms associated to the gauge fields $b_{I\mu}$, since they do not essentially contribute in the IR.}, 
\begin{eqnarray}
	\label{Eq:L-dual-ep}
	\widetilde{\mathcal{L}}&=&\sum_{I=1,2}\left[|(\partial_\mu-ib_{I\mu})\phi_I|^2+m^2|\phi_I|^2+\frac{\lambda}{2}|\phi_I|^4\right]
	\nonumber\\
	&+&\frac{M^2}{2}(b_{1\mu}+
	b_{2\mu})^2.
\end{eqnarray}
The gauge invariance of the above dual theory is more restrictive than the one of Eq. (\ref{Eq:Dual-L}). Indeed, while for the latter independent gauge transformations, $b_{I\mu}\to b_{I\mu}+\partial_\mu\xi_I$, $\phi_I\to e^{i\xi_I}\phi_I$, leave the action invariant, we have to use $b_{1\mu}\to b_{1\mu}+\partial_\mu\xi$, $b_{2\mu}\to b_{2\mu}-\partial_\mu\xi$, $\phi_1\to e^{i\xi}\phi_1$,  $\phi_2\to e^{-i\xi}\phi_2$ in Eq. (\ref{Eq:L-dual-ep}). Closely related to this is the fact that effectively $M^2$ is very large, since it is proportional to $e^2$; recall the closely related discussion in this context in Sect. \ref{Sec:CP-PV}. This leads effectively to only one gauge field, ultimately implying that the theory is self-dual \cite{Motrunich-Vishwanath_PhysRevB.70.075104,Senthil_PhysRevB.70.144407}. On the contrary, there is nothing similarly happening to the mixed CS term arising in Eq. (\ref{Eq:Dual-L}). The two gauge fields appearing there are a robust feature of the theory, which is also reflected in its broader gauge freedom. There is no need to restrict $\xi_2=-\xi_1$ in this case.  

The effective reduction to a single gauge field also occurs in the bosonization duality of the theory leading to the bosonic dual Lagrangian (\ref{Eq:L-dual-ep}). This contrasts to the fermionic dual theory of Eq. (\ref{Eq:fAction}). This fact has far reaching consequences in the study of possible critical points \cite{Senthil_PhysRevX.7.031051,Shyta_PhysRevLett.127.045701}. One major difficulty in this case is whether or not the system becomes critical and this has been a matter of some controversy in the field of deconfined quantum critical points. Numerical results typically yield a first-order phase transition for the standard (i.e., without a CS term) $U(1)\times U(1)$ Abelian Higgs model \cite{KUKLOV20061602,Kragset_PhysRevLett.97.247201}, so quantum criticality would be absent. Nevertheless, as pointed out recently in Ref. \cite{Senthil_PhysRevX.7.031051}, the bosonization duality rather suggests that the theory becomes critical. The dual fermion model in this case is given by the Lagrangian, 
\begin{equation}
	\label{Eq:QED-2-fermions}
	\mathcal{L}_f=\sum_{I=1,2}\bar{\psi}_I(\slashchar{\partial}-i\slashchar{a})\psi_I,
\end{equation}
where we have omitted the BF terms. At first sight it seems obvious that this theory is critical, with a current correlator being given by twice the expression of Eq. (\ref{Eq:CC-F-massless}) of Appendix \ref{app:free-fermions}, since the above theory features two massless fermions. This result is the same as the one-loop vacuum polarization at the critical point (meaning $e^2\to\infty$). This corresponds to an IR stable fixed point in terms of a properly defined dimensionless gauge coupling. Higher loops do not seem to change this picture. However, non-perturbative behavior may spoil the criticality, causing a dynamical generation of a fermion mass \cite{Appelquist_PhysRevLett.60.2575,Rantner-Wen_PhysRevLett.86.3871,Franz_PhysRevB.66.054535,Herbut_PhysRevLett.97.146401,Herbut_PhysRevB.66.094504,Gusynin_PhysRevB.67.115201}. This is best seen by generalizing the Lagrangian of Eq. (\ref{Eq:QED-2-fermions}) to include $2N$ fermion components rather than two. Grouping the $2N$ two-component Dirac spinors $\psi_I$ into $N$ four-component Dirac spinors $\Psi_a$, we can introduce $4\times 4$ gamma matrices that also satisfy a Clifford algebra, along with a  $\gamma_5$-like matrix \cite{Appelquist_PhysRevD.33.3704}, introducing a chiral symmetry reminiscent of the actual chiral symmetry for massless fermions in even dimensions of spacetime \footnote{Putting the fermions in the lattice, this chiral symmetry actually manifests itself as a sublattice symmetry. Graphene constitutes a realization of this type of chiral symmetry; see, for instance, Ref. \cite{Herbut_PhysRevLett.97.146401}.}. A fermion QED in 2+1 dimensions defined in this way does not break parity if a mass term of the form $m\sum_{a=1}^N\bar{\Psi}_a\Psi_a$ is added, since in terms of the original two-component fermions it corresponds to having half of the fermions with mass $m$ and the other half with mass $-m$. Hence, there is no anti-symmetric contribution to the vacuum polarization in this case and no CS term is generated.  Since $e^2\to\infty$, the gauge field propagator is obtained by inverting the vacuum polarization. The Schwinger-Dyson equation for the fermion propagator then is known to imply a dynamically generated mass that vanishes for $N\geq 32/\pi^2\approx 3.24$, a result following from an approximate self-consistent solution \cite{Appelquist_PhysRevLett.60.2575}. Thus, if one assumes the validity of this non-perturbative result \footnote{Strictly speaking, this result is derived by first assuming a large $N$ limit and neglecting vertex corrections in the Schwinger-Dyson equation. We could in principle question a critical value of $N$ that is not even large. However, numerical analyses using staggered fermions in the lattice seem to corroborate the existence of a critical value of $N$ that is not large \cite{Strouthos_2009}; see Ref. \cite{Braun_PhysRevD.90.036002} for a recent account of this problem and references therein.}, we obtain that the theory of Eq. (\ref{Eq:QED-2-fermions}) becomes gapped once non-perturbative quantum fluctuation effects are accounted for. However, if this is the case, it would be difficult to even claim the existence of a phase transition at all for $N=2$. On the other hand, there is some evidence favoring a critical value of $N$ lying in the interval $1<N<2$ \cite{Braun_PhysRevD.90.036002, Appelquist_PhysRevD.60.045003, Gusynin_PhysRevD.94.125009}. In this case, we can safely argue that the bosonization duality rather implies a quantum critical point for the $U(1)\times U(1)$ symmetric Abelian Higgs model. However, recent conformal bootstrap calculations \cite{li2018solving, li2021conformality, albayrak2022bootstrapping} show that the theory is gapped for $N=1$ (or two two-component fermions), so the theory is not critical. 

It is unclear at the moment whether such an IR form \eqref{Eq:QED-2-fermions} of the dual theory can  actually capture the actual phase structure of the lattice model, which in this case corresponds to a quantum antiferromagnet with easy-plane anisotropy \cite{Motrunich-Vishwanath_PhysRevB.70.075104, motrunich2008comparative, shyta2021frozen}. Furthermore, so far simulations address rather the so called deep easy-plane limit, so the phase stiffnesses are the same. On the other hand, there is numerical evidence for a second-order phase transition when the phase stiffnesses are different \cite{Sudbo_PhysRevB.82.134511}. 

The difficulties above are not shared by the $U(1)\times U(1)$ symmetric Abelian Higgs model with a CS term considered recently in Ref. \cite{Shyta_PhysRevLett.127.045701} and analyzed in Sec. \ref{Sec:BosDual2-comp}. Importantly, the fermionic dual theory given in Eq. (\ref{Eq:fAction}) features two gauge fields and a reduction to a single gauge field does not occur. This theory features also dynamical CS terms, including mixed ones \footnote{If background fields are included, CS terms involving only background fields also appear in the theory of Eq. (\ref{Eq:QED-2-fermions}) \cite{Senthil_PhysRevX.7.031051}. This does not affect the critical behavior, however}. A one-loop RG analysis of the original model (\ref{eq:CP1soft}) shows that a critical point indeed exists \cite{Shyta_PhysRevLett.127.045701}. Most importantly, the fermionic dual action does not lead to dynamical mass generation. To see this, we consider the Schwinger-Dyson equation for one of the fermions $\psi_I$, 
\begin{equation}
	\label{Eq:SD}
	G^{-1}(p)=i\gamma\cdot p+\int\frac{d^3q}{(2\pi)^3}\gamma_\mu G(q)\gamma_\nu D_{\mu\nu}(p-q),
\end{equation}
where $D_{\mu\nu}(p)$ is the propagator of Eq. (\ref{Eq:D-b1b1}) and $G(p)=[Z(p)i\gamma\cdot p+\Sigma(p)]^{-1}$ represents the full fermionic propagator. In the above equation vertex corrections are being neglected. In order to gain better control of the calculation, we introduce $M$ copies of fermions $\psi_1$ and $\psi_2$, and thus we have to multiply the diagonal matrix elements in Eq. (\ref{eq:matrixM}) by $M$. This modifies $D_{\mu\nu}(p)$ to, 
\begin{equation}
	D_{\mu\nu}(p)=\frac{16\pi^2M}{(64+M^2\pi^2)|p|}\left(\delta_{\mu \nu}-\frac{p_\mu p_\nu}{p^2}\right). 
\end{equation}
Equation (\ref{Eq:SD}) leads then to two self-consistent integral equations of the form,
\begin{widetext}
\begin{equation} 
	\label{eq:SigmaQEDeq}
	\Sigma(p)=\frac{32\pi^2M}{64+M^2\pi^2}\int\frac{d^3k}{(2\pi)^3}\frac{\Sigma(k)}
	{[Z^2(k)k^2+\Sigma^2(k)]|k+p|},
\end{equation}
\begin{equation}
	\label{eq:ZQEDeq}
	Z(p)=1-\frac{16\pi^2M}{(64+M^2\pi^2)p^2}\int\frac{d^3k}{(2\pi)^3}\frac{[k^2-p^2+(k+p)^2](k+p)\cdot p 
		Z(k)}{[Z^2(k)k^2+\Sigma^2(k)]|k+p|^3}.
\end{equation}	
\end{widetext}
Using a well-known procedure \cite{Appelquist_PhysRevD.33.3704,Appelquist_PhysRevLett.60.2575,Nash_PhysRevLett.62.3024}, we easily convert Eqs. (\ref{eq:SigmaQEDeq}) and (\ref{eq:ZQEDeq}) to nonlinear differential equations, 
\begin{equation}
	\label{Eq:d-Sig}
	\frac{d}{ds}\left[s^2\frac{d\Sigma(s)}{ds}\right]=
	-\frac{16M}{64+M^2\pi^2}\frac{s^2\Sigma(s)}{Z^2(s)s^2+\Sigma^2(s)},
\end{equation}
\begin{equation}
	\label{Eq:d-Z}
	\frac{d}{ds}\left[s^4\frac{dZ(s)}{ds}\right]=\frac{16M}{64+M^2\pi^2}
	\frac{s^4Z(s)}{Z^2(s)s^2+\Sigma^2(s)},
\end{equation}
where $s\equiv|p|$. These equations are solved with the boundary conditions, $\lim_{s\to 0}s\Sigma(s)=0$, $sd\Sigma(s)/ds|_{s=\Lambda}=-\Sigma(\Lambda)$, $sdZ(s)/ds|_{s=\Lambda}=3[1-Z(\Lambda)]$, where $\Lambda$ is a UV cutoff. In addition, the spectral representation implies that $0<Z(0)\leq 1$ must hold. Typically a first approximation to solve Eqs. (\ref{Eq:d-Sig}) and (\ref{Eq:d-Z}) is by linearizing them. Using the leading order result where $Z(0)=1$, we obtain a nonzero gap of the form, 
\begin{equation}
	\label{Eq:Sigma_0}
	\Sigma_\pm(0)=\pm\Lambda e^{-2\pi/\gamma},
\end{equation}
where, 
\begin{equation}
	\label{Eq:gamma}
	\gamma=\sqrt{\frac{64M}{64+\pi^2M^2}-1}.
\end{equation}
We have included two possible signs for the generated gap, since we are dealing with two-component spinors here, so strictly speaking a chiral symmetry breaking is only possible for $M=2N$, with half of the fermions getting a mass $\Sigma_+(0)$ and the other half $\Sigma_-(0)$. For $M=2N+1$ it is the parity that is being spontaneously broken. 
  
We obtain that $\Sigma(0)/\Lambda$ is only nonzero for $M_-<M<M_+$, where $M_\pm=(64\pm\sqrt{4096-256\pi^2})/(2\pi^2)$ are the critical values of $N$ for which the gap vanishes. We note that $M_-\approx 1.235$, so this solution does not yield any gap generation for $M=1$. Even for the case where $M$ lies in the interval $(M_-,M_+)$, the corresponding value of $\Sigma(0)/\Lambda$ is tiny. Thus, for all practical purposes we can assume that there is no gap generation at all. This situation is in sharp contrast with the one corresponding to the Lagrangian of Eq. (\ref{Eq:QED-2-fermions}) when it is generalized to include $2N$ fermion fields. Of course, one important difference is that in the case just discussed $M$ fermions are coupled to the gauge field $b_{1\mu}$ while the other $M$ fermions couple to the gauge field $b_{2\mu}$, but this is actually not the main point. The fact is that applying the same Schwinger-Dyson method to Eq. (\ref{Eq:QED-2-fermions}), would amount to replacing $64M/(64+\pi^2M^2)$ in Eq. (\ref{Eq:gamma}) by $32/(\pi^2N)$, leading in this way to the already quoted result that the gap vanishes only for $N\geq 32/\pi^2\approx 3.24$. Note that the number $f=32/(\pi^2N)$ is proportional to the dimensionless gauge coupling at the IR stable fixed point (to see this we have to include a Maxwell term as a UV regulator, similarly to the discussion in Sect. \ref{Sec:PV-duality}) \cite{Appelquist_PhysRevD.33.3704}, and thus  $64M/(64+\pi^2M^2)$ with $M=2N$ would coincide with $f$ only in the large $N$ limit.   

 We have thus found that only values of $M=2N>1$ would allow for chiral symmetry breaking. We note however that the interval $(M_-,M_+)$ includes the values 2, 3, and 4. Among those values, $M=3$ is special, since it is odd and this would mean that parity gets dynamically broken in this case. Such a result is not expected in ordinary QED in 2+1 dimensions \cite{Appelquist_PhysRevD.33.3774} and seems to violate a theorem by Vafa and Witten \cite{Vafa-Witten}, which states that in massless QED in (2+1)$d$ it is not possible to spontaneously break parity for any value of $M$. However, in our theory the dependence on $M$ arises in a somewhat different way, since the photon propagator is obtained from inverting the matrix appearing in Eq. (\ref{eq:matrixM}). Thus, the conditions underlying the Vafa-Witten result \cite{Vafa-Witten} and the early Schwinger-Dyson calculation performed in Ref. \cite{Appelquist_PhysRevD.33.3774} do not apply in this case. Furthermore, for the $M=1$ case, which is our main interest here, no dynamical mass generation occurs. Hence, the critical current correlator  obtained in Sect. \ref{Sec:BosDual2-comp} should correctly account for the scale invariance of the theory.              

\section{Conclusion}
\label{Sec:Conclusion}


Let us summarize the main results of our paper. The first important result we have obtained is the bosonization duality for the topological $U(1)\times U(1)$ Abelian Higgs model, which in terms of Lagrangians can be schematically represented in the form, 
\begin{eqnarray}
	\label{Eq:MainResult}
	\mathcal{L}_b&=&\sum_{a=1,2}|(\partial_\mu-ia_\mu-i(-1)^aB_\mu)z_a|^2
	\nonumber\\
	&+&\frac{i}{4\pi}\epsilon_{\mu\nu\lambda}a_\mu\partial_\nu a_\lambda+\dots
	\nonumber\\
	&&\hspace{2cm} \big\Updownarrow
	\nonumber\\
	\mathcal{L}_f&=&\sum_{I=1,2}\left[ \bar{\psi}_I(\slashchar{\partial}-i\slashchar{b}_I)\psi_I+\frac{i}{8\pi}b_{I\mu}\epsilon_{\mu\nu\lambda}\partial_\nu b_{I\lambda}\right]
	\nonumber\\
	&+& \frac{i}{2\pi}b_{1\mu}\epsilon_{\mu\nu\lambda}\partial_\nu b_{2\lambda}+\frac{i}{2\pi}\epsilon_{\mu \nu \lambda}B_\mu\partial_\nu (b_{2\lambda}-b_{1\lambda}),
	\nonumber\\
\end{eqnarray}
where $B_\mu$ is a background field. While paving the way to derive the main duality (\ref{Eq:MainResult}), we have revisited some of the known results of the duality web with focus on inferring critical properties and current correlators. Note that in contrast with the duality of Eq. (\ref{Eq:Boson+flux}) for the $U(1)$-invariant case (i.e., featuring a single scalar), the fermionic dual model is not a free theory in the case of Eq. (\ref{Eq:MainResult}). We have demonstrated for the duality (\ref{Eq:MainResult}) that the topological and Noether current correlators only differ by a universal prefactor we have calculated, giving the result of Eq. (\ref{Eq:Kf=Cb-1}). The agreement of bosonic and fermionic current correlation functions in the bosonization duality also highlights that theories at criticality flow to the same fixed points and, therefore, exhibit similar critical behavior. Thus, an important feature of the topological $U(1)\times U(1)$ Abelian Higgs model is that the system undergoes a second-order phase transition, in contrast to the usual non-topological case, where the existence of the critical point is still under debate. Indeed, the Schwinger-Dyson analysis of the fermionic dual model appearing in Eq. (\ref{Eq:MainResult}) complements and strengthens our RG analysis of the bosonic Lagrangian (\ref{eq:CP1soft}) [represented in a more schematic form in Eq. (\ref{Eq:MainResult})] in Ref. \cite{Shyta_PhysRevLett.127.045701}, since it shows that no dynamical mass generation occurs. As was extensively discussed in the text, the bosonization duality for the non-topological $U(1)\times U(1)$ theory, despite having a much simpler looking form, yields far less conclusive results regarding its phase structure.

\begin{acknowledgments}
	We thank the Deutsche Forschungsgemeinschaft (DFG) 
	for support through the W\"urzburg-Dresden Cluster of Excellence on Complexity and Topology in Quantum Matter – ct.qmat (EXC 2147, project-id 39085490) and 
	the Collaborative Research Center  
	SFB 1143 (project-id 247310070). 
\end{acknowledgments}

 \appendix
 \section{Current correlator in the $O(2)$ nonlinear sigma model}
 \label{app:O(2)-nlsm}
 
 If we consider a scalar massless theory with an additional constraint, $|\varphi|^2=1$, the model turns into an $O(2)$ nonlinear $\sigma$ model. In this case the constraint 
 is easily solved by $\varphi=e^{i\theta}$, and the Lagrangian reduces to, 
 \begin{equation}
 	\label{Eq:Free-scalar}
 	\mathcal{L}=\frac{1}{2g}(\partial_\mu\theta)^2,
 \end{equation} 
 where we have included now a coupling constant $g$. 
 This is the model for a free scalar field theory. However, since $\theta$ is periodic, such 
 a model is nontrivial. For instance, when supplemented by a coupling constant, it is known 
 to lead to a Berezinskii-Kosterlitz-Thouless (BKT) phase transition at $d=2$. 
 
 The conserved current associated to the Lagrangian (\ref{Eq:Free-scalar}) is given by, 
 \begin{equation}
 	\label{eq:Jmu}
 	J_\mu=\frac{1}{g}\partial_\mu\theta,
 \end{equation}
 and makes its appearance in Sec. \ref{sec:1plus1Dirac} in the context of a bosonized version of massless Dirac fermions. 
 The corresponding  current correlation function is given by, 
 \begin{equation}
 	K_{\mu\nu}(x-x')=\frac{1}{g^2}\langle\partial_\mu\theta(x)\partial_\nu'\theta(x')\rangle. 
 \end{equation}
 This can be rewritten as, 
 \begin{equation}
 	K_{\mu\nu}(x)=-\frac{1}{g^2}\partial_\mu\partial_\nu\langle\theta(x)\theta(0)\rangle. 
 \end{equation}
 Since, 
 \begin{equation}
 	\langle\theta(x)\theta(0)\rangle=\frac{g}{S_d(d-2)|x|^{d-2}},
 \end{equation}
 we easily obtain that, 
 \begin{equation}
 	\label{Eq:K}
 	K_{\mu\nu}(x)=\frac{1}{gS_d|x|^d}\left(\delta_{\mu\nu}-d\frac{x_\mu x_\nu}{x^2}\right),
 \end{equation}
 and we also verify that this result is transverse, 
 \begin{equation}
 	\partial_\mu K_{\mu\nu}=0.
 \end{equation}
 However, the current correlation function is not transverse in  momentum space, 
 %
 \begin{equation}
 	K_{\mu\nu}(p)=\frac{p_\mu p_\nu}{gp^2}.
 \end{equation}
 This is reminiscent of the axial anomaly occurring naturally in even spacetime dimensions. This behavior has another consequence: the trace 
 $K_{\mu\mu}(x)$ vanishes, but the same does not happen for $K_{\mu\mu}(p)$, which yields the  unity. In fact, we actually have, 
 \begin{equation}
 	K_{\mu\mu}(x)=\frac{1}{g}\delta^d(x),
 \end{equation} 
 i.e., $K_{\mu\mu}(x)$ indeed vanishes everywhere, except at $x=0$ where a delta function 
 singularity arises. 
 
 On the other hand, we can recall that there is another conserved current, an axial one, namely \cite{zinn2002quantum},
 \begin{equation}
 	\tilde{J}_\mu=\frac{1}{g}\epsilon_{\mu\nu}\partial_\nu\theta,
 \end{equation} 
 which is dual to $J_\mu$ of Eq. \eqref{eq:Jmu}. The current correlation function $\widetilde{K}_{\mu\nu}(x)=\langle\tilde{J}_\mu(x)\tilde{J_\nu}(0)\rangle$ is simply given by, 
 \begin{equation}
 	\widetilde{K}_{\mu\nu}(x)=-K_{\mu\nu}(x),
 \end{equation}
 implying they are both transverse in real space. However, as we have already seen, $K_{\mu\nu}(p)$ is longitudinal. The interesting result following from the simple calculation above is that, $\widetilde{K}_{\mu\nu}(p)$, by contrast, is transverse, 
 \begin{equation}
 	\widetilde{K}_{\mu\nu}(p)=\frac{1}{g}\left(\delta_{\mu\nu}-\frac{p_\mu p_\nu}{p^2}\right). 
 \end{equation}

 \section{Current correlation function of free theories}
 \label{app:ints}
 
 \subsection{Free scalar bosons in $d$ dimensions}
 
 A general expression for the bosonic current correlation function of the  massless theory in $d$ dimensions can be obtained explicitly. 
 %
 The vacuum polarization in scalar QED in the massless limit is easily shown to precisely correspond to the exact current correlation function of the free theory, 
 \begin{equation}
 	\label{Eq:L-free}
 	\mathcal{L}=|\partial_\mu\varphi|^2+m^2|\varphi|^2,
 \end{equation}
 when $m=0$. The theory possesses a conserved $U(1)$ current given by,
 \begin{equation}
 	j_\mu=-i(\varphi^*\partial_\mu\varphi-\varphi\partial_\mu\varphi^*).
 \end{equation}
 The corresponding current correlation function for a free complex scalar boson can be calculated explicitly and initially is written as, 
 \begin{widetext}
 	\begin{eqnarray}
 		C_{\mu\nu}(x-x')&=&\langle j_\mu(x)j_\nu(x')\rangle=-\langle\varphi^*(x)\partial_\mu\varphi(x)\varphi^*(x')\partial_\nu'\varphi(x')
 		\rangle-\langle\varphi(x)\partial_\mu\varphi^*(x)\varphi(x')\partial_\nu'\varphi^*(x')
 		\rangle\nonumber\\
 		&+&\langle\varphi^*(x)\partial_\mu\varphi(x)\varphi(x')\partial_\nu'\varphi^*(x')
 		\rangle+\langle\varphi(x)\partial_\mu\varphi^*(x)\varphi^*(x')\partial_\nu'\varphi(x')
 		\rangle.
 	\end{eqnarray}	
 	
 		Since the theory is free, we can use Wick's theorem to decouple the correlations above. The 
 		result is, 
 		%
 		\begin{eqnarray}
 			\label{Eq:C}
 			C_{\mu\nu}(x-x')&=&-\langle\varphi^*(x)\partial_\nu'\varphi(x')\rangle\langle\partial_\mu
 			\varphi(x)\varphi^*(x')\rangle-\langle\varphi(x)\partial_\nu'\varphi^*(x')\rangle\langle\partial_\mu
 			\varphi^*(x)\varphi(x')\rangle\nonumber\\
 			&+&\langle\varphi^*(x)\varphi(x')\rangle\langle\partial_\mu\varphi(x)\partial_\nu'
 			\varphi^*(x')\rangle+\langle\varphi(x)\varphi^*(x')\rangle\langle\partial_\mu\varphi^*(x)\partial_\nu'
 			\varphi(x')\rangle.
 		\end{eqnarray}
 	\end{widetext}
 	We note the Fourier representations of the correlation functions, 
 	\begin{equation}
 		\label{Eq:C1}
 		\langle\varphi(x)\varphi^*(x')\rangle=\int_p\frac{e^{ip\cdot(x-x')}}{p^2+m^2},
 	\end{equation}
 	\begin{equation}
 		\label{Eq:C2}
 		\langle\partial_\mu\varphi(x)\varphi^*(x')\rangle=i\int_p\frac{p_\mu e^{ip\cdot(x-x')}}{p^2+m^2},
 	\end{equation}
 	\begin{equation}
 		\label{Eq:C3}
 		\langle\partial_\mu\varphi(x)\partial_\nu'
 		\varphi^*(x')\rangle=\int_p\frac{p_\mu p_\nu e^{ip\cdot(x-x')}}{p^2+m^2},
 	\end{equation}
 	while the other correlation functions entering Eq. (\ref{Eq:C}) follow by complex 
 	conjugation. In the expressions above, we used the shorthand notation, 
 	\begin{equation}
 		\int_k=\int\frac{d^dk}{(2\pi)^3}.
 	\end{equation} 
 	After inserting the Fourier representations into Eq. (\ref{Eq:C}), we obtain 
 	the Fourier transform of the current correlation function as, 
 	\begin{equation}
 		C_{\mu\nu}(k)=\int d^3xe^{-ik\cdot x}C_{\mu\nu}(x).
 	\end{equation}
 	Finally, some straightforward algebraic manipulations and simplifications yield, 
 	\begin{equation}
 		\label{Eq:C-I-rel}
 		C_{\mu\nu}(k)=I_{\mu\nu}(k),
 	\end{equation}
 	where $I_{\mu\nu}(k)$ is given in momentum space by the integral, 
 	\begin{equation}
 		\label{Eq:Imunu}
 		I_{\mu\nu}(q)=\int_k\frac{(2k-q)_\mu(2k-q)_\nu}{[(k-q)^2+m^2](k^2+m^2)}.
 	\end{equation}
 	Since we can write, 
 	\begin{equation}
 		I_{\mu\nu}(q)=A(q)\delta_{\mu\nu}+B(q)q_\mu q_\nu, 
 	\end{equation}
 	where, 
 	\begin{equation}
 		A(q)=\frac{1}{d-1}\left[I_{\mu\mu}(q)-\frac{q_\mu q_\nu}{q^2}I_{\mu\nu}(q)\right],
 	\end{equation}
 	\begin{equation}
 		B(q)=\frac{1}{(1-d)q^2}\left[I_{\mu\mu}(q)-d\frac{q_\mu q_\nu}{q^2}I_{\mu\nu}(q)\right],
 	\end{equation}
 	we obtain after straightforward simplifications,
 	\begin{eqnarray}
 		\label{Eq:A}
 		A(q)&=&-\frac{1}{d-1}\left\{\frac{4m^2}{d-2}\int_k\frac{1}{(k^2+m^2)^2}
 		\right.\\
 		&+&\left.(q^2+4m^2)\int_k\frac{1}{[(k-q)^2+m^2](k^2+m^2)}
 		\right\}.
 		\nonumber
 	\end{eqnarray}
 	\begin{eqnarray}
 		\label{Eq:B}
 		B(q)&=&\frac{1}{(d-1)q^2}\left\{-4m^2\int_k\frac{1}{(k^2+m^2)^2}
 		\right.\\
 		&+&\left.(q^2+4m^2)\int_k\frac{1}{[(k-q)^2+m^2](k^2+m^2)}
 		\right\}.\nonumber
 	\end{eqnarray}
 	
 	We now decompose $I_{\mu\nu}(q)$ into transverse and longitudinal components, 
 	\begin{equation}
 		I_{\mu\nu}(q)=A(q)\left(\delta_{\mu\nu}-\frac{q_\mu q_\nu}{q^2}\right)
 		+\left[B(q)q^2+A(q)\right]\frac{q_\mu q_\nu}{q^2}.
 	\end{equation}
 	The longitudinal component is given simply by, 
 	\begin{equation}
 		B(q)q^2+A(q)=-\frac{4m^2}{d-2}\int_k\frac{1}{(k^2+m^2)^2},
 	\end{equation}
 	and we see that for $m^2=0$ the result for $I_{\mu\nu}(q)$ is purely transverse, 
 	\begin{equation}
 		\label{Eq:I}
 		I_{\mu\nu}(q)|_{m^2=0}=-\Pi(q)(q^2\delta_{\mu\nu}-q_\mu q_\nu),
 	\end{equation}
 	with
 	\begin{equation}
 		\label{Eq:Pi-massless}
 		\Pi(q)=\int_k\frac{1}{(k-q)^2k^2}=c_d|q|^{d-4},
 	\end{equation}
 	where $c_d$ is given by,
 	\begin{equation}
 		\label{Eq:cd}
 		c_d=\frac{d-2}{(4\pi)^{d/2}}\Gamma\left(
 		2-\frac{d}{2}\right)\frac{\Gamma^2(d/2-1)}{\Gamma(d)}.
 	\end{equation}
 	
 	Let us now specialize to $d=3$. In this case we have, 
 	\begin{equation}
 		\int_k\frac{1}{(k^2+m^2)^2}=\frac{1}{8\pi|m|},
 	\end{equation}
 	\begin{eqnarray}
 		\int_k\frac{1}{[(k-q)^2+m^2](k^2+m^2)}=\frac{1}{4\pi|q|}\arctan\left(
 		\frac{|q|}{2|m|}\right).
 		\nonumber\\
 	\end{eqnarray}
 	Hence, the transverse and longitudinal components of $I_{\mu\nu}(q)$ read, 
 	\begin{equation}
 		I_t(q)=-\frac{1}{4\pi}\left[|m|+\frac{(q^2+4m^2)}{2|q|}\arctan\left(
 		\frac{|q|}{2|m|}\right)\right],
 	\end{equation}
 	\begin{equation}
 		I_l(q)=-\frac{|m|}{2\pi}.
 	\end{equation}
 	In the expression for $A(q)$ we have used dimensional regularization to write,
 	\begin{equation}
 		\label{Eq:dimregrel}
 		\int_k\frac{1}{k^2+m^2}=-\frac{2m^2}{d-2}\int_k\frac{1}{(k^2+m^2)^2},
 	\end{equation}
 	which removes a power of the UV cutoff $\Lambda$ of the form $\sim\Lambda^{d-2}$ spoiling gauge invariance. 
 	
 	Therefore, the current correlation function of a massless bosonic theory is purely transverse and we obtain in real space that, 
 	\begin{equation}
 		\label{Eq:C-1}
 		C_{\mu\nu}(x)=\frac{2}{S_d^2(d-2)|x|^{2(d-1)}}\left(\delta_{\mu\nu}-2\frac{x_\mu x_\nu}{x^2}\right),
 	\end{equation} 
 	where $S_d=2\pi^{d/2}/\Gamma(d/2)$ is the solid angle in $d$ dimensions.  
 	Indeed, we verify that $\partial_\mu C_{\mu\nu}(x)=0$, and in momentum space we have, 
 	\begin{equation}
 		\label{Eq:C-1-p}
 		C_{\mu\nu}(q)=-c_d|q|^{d-2}\left(\delta_{\mu\nu}-\frac{q_\mu q_\nu}{q^2}\right).
 	\end{equation}
 	\subsection{Free fermions in 2+1 dimensions} 
 	\label{app:free-fermions}
 	
 	It is a straightforward exercise to exactly compute the current correlation function for free fermions. In momentum 
 	space we have, 
 	\begin{equation}
 		\label{Eq:fermion-cc-correl}
 		C_{\mu\nu}^F(p)=\langle j_\mu(p)j_\nu(-p)\rangle=-\int_q{\rm tr}[\gamma_\mu G(q)\gamma_\nu G(p-q)],
 	\end{equation} 
 	where, 
 	\begin{equation}
 		G(p)=\frac{M-i\slashchar{p}}{p^2+M^2}.
 	\end{equation}
 	Thus, 
 	\begin{equation}
 		C_{\mu\nu}^F(p)=\int_q\frac{{\rm tr}[\gamma_\mu(M-i\slashchar{q})\gamma_\nu(M-i\slashchar{p}+i\slashchar{q})]}
 		{(q^2+M^2)[(p-q)^2+M^2]}.
 	\end{equation}
 	The trace in the numerator of the above expression is rewritten as, 
 	\begin{eqnarray}
 		&&{\rm tr}[\gamma_\mu(M-i\slashchar{q})\gamma_\nu(M-i\slashchar{p}+i\slashchar{q})]=
 		2(M^2+q^2\nonumber\\
 		&-&p\cdot q)\delta_{\mu\nu}-2(p_\mu q_\nu+p_\nu q_\mu)+4 q_\mu q_\nu 
 		\nonumber\\
 		&+&2M\epsilon_{\mu\nu\lambda}p_\lambda,
 	\end{eqnarray}
 	where we have made use of the identities, 
 	\begin{equation}
 		{\rm tr}(\gamma_\mu\gamma_\nu)=2\delta_{\mu\nu},
 	\end{equation}
 	\begin{equation}
 		{\rm tr}(\gamma_\mu\gamma_\nu\gamma_\lambda)=2i\epsilon_{\mu\nu\lambda},
 	\end{equation}
 	\begin{equation}
 		{\rm tr}(\gamma_\mu\gamma_\lambda\gamma_\nu\gamma_\rho)=2(\delta_{\mu\lambda}
 		\delta_{\nu\rho}+\delta_{\mu\rho}\delta_{\lambda\nu}-\delta_{\mu\nu}\delta_{\lambda\rho}).
 	\end{equation}
 	Performing explicitly the several integrals yields,
 	\begin{eqnarray}
 		\label{Eq:CC-F}
 		&&-C_{\mu\nu}^F(p)=\frac{1}{4\pi}\left[|M|+\frac{p^2-4M^2}{2|p|}\arctan\left(
 		\frac{|p|}{2|M|}\right)\right]\nonumber\\
 		&\times&\left(\delta_{\mu\nu}-\frac{p_\mu p_\nu}{p^2}\right)
 		+\frac{M}{2\pi|p|}\arctan\left(\frac{|p|}{2|M|}\right)\epsilon_{\mu\nu\lambda}p_\lambda,
 		\nonumber\\
 	\end{eqnarray}
 	where UV finiteness of the parity-even transverse term has been obtained via dimensional regularization. We will see later that a different gauge-invariant regularization is more appropriate in this case. 
 	
 	For $M=0$ the above result becomes simply, 
 	\begin{equation}
 		\label{Eq:CC-F-massless}
 		C_{\mu\nu}^F(p)|_{M=0}=-\frac{|p|}{16}\left(\delta_{\mu\nu}-\frac{p_\mu p_\nu}{p^2}\right). 
 	\end{equation}
 	Up to a minus sign, Eq. (\ref{Eq:CC-F-massless}) coincides with the bosonic current correlation function for free massless bosons of Eq. (\ref{Eq:C-1-p}) when $d=3$ is set,  
 	\begin{equation}
 		C_{\mu\nu}^F(p)|_{M=0}=C_{\mu\nu}(p)|_{m=0}.
 	\end{equation}

\bibliography{bosdualcurrent}

\end{document}